\documentclass[prl,twocolumn,amssymb,amsmath,aps,showpacs,floatfix,reprint]{revtex4-2}
\usepackage[colorlinks=true,linkcolor=blue,citecolor=blue,urlcolor=blue]{hyperref}
\usepackage{graphicx,bm}
\usepackage{xcolor}

\begin{document}

\title{Rotational viscosity in spin resonance of hydrodynamic electrons}

\author{ A.~N.~Afanasiev}
\email{afanasiev.an@mail.ru}
\author{P.~S.~Alekseev}
\author{A.~A.~Danilenko}
\author{A.~A.~Greshnov}
\author{M.~A.~Semina}
\affiliation{Ioffe Institute, St.~Petersburg 194021, Russia}

\begin{abstract}

In novel ultra-pure materials electrons can form a viscous fluid, which is fundamentally different by its dynamics from the electron gas in ordinary  conductors with significant density of defects. The shape of the non-stationary flow of such electron fluid  is similar to the alternating flow of blood in large-radius arteries~[J.~R.~Womersley, \href{https://physoc.onlinelibrary.wiley.com/doi/abs/10.1113/jphysiol.1955.sp005276}{J. Physiol. {\bf 127}, 552 (1955)}]. The rotational viscosity effect  is responsible for interconnection between the dynamics of electron spins and flow inhomogeneities. In particular, it induces the spin polarization of electrons in a curled flow via an internal spin-orbit torque acting on electron spins. Here we show that this effect in an electron fluid placed in a magnetic field leads  to a correction to the ac sample impedance, which has a resonance at the  Larmor frequency of electrons. In this way, \textit{via} the electrically detected spin resonance the Womersley flow of an electron fluid can be visualized and the rotational viscosity can be measured.


\end{abstract}

\maketitle


{\em 1. Introduction.} In ultra-pure materials with small density of defects, conduction electrons can form a viscous fluid at low temperatures. The electric transport in such fluid occurs via the formation of inhomogeneous hydrodynamic flows, controlled by particular shapes of samples. This idea was put forward many years ago for 3D metals with strong electron-phonon coupling~\cite{Gurzhi}. Recently, the hydrodynamic regime of electron transport has been realized in high-quality samples of graphene~\cite{grahene, grahene_2, grahene_3, Levitov_et_al, profile_1, profile_2}, quasi-two dimensional metal PdCoO$_2$~\cite{Weyl_sem_1}, Weyl semimetal WP$_2$~\cite{Weyl_sem_2}, and high-mobility GaAs quantum wells~\cite{exps_neg_1, exps_neg_2, exps_neg_3,exps_neg_4,je_visc,Gusev_1,Gusev_2,Gusev_3,recent___1,recent___2}. These experiments induced an avalanche of theoretical works (see, for example Refs.~\cite{ph_tr_num, Levitov_et_al_2, Lucas, eta_xy, we_3, Lucas_2, we_4, we_5_1, we_5_2, we_5, we_6, recentest_, recentest___breathing_flow, recentest2, L_n_1, recentest3, vis_res, ph_tr_num__ball_formulas, Khoo_Villadiego, future, Alekseev_Alekseeva, Alekseev_Dmitriev, future2, we_7, we_8___ph_tr_formulas}), which were aimed to formulation and search for the evidences of the hydrodynamic regime as well as to studying various types and regimes of flows of the electron fluid.

An electric current in samples with a noticeable spin-orbit interaction induces various interesting spin-dependent transport effects. It was predicted nearly half a century ago for Ohmic conductors, where the scattering of electrons on disorder dominates, that the spin-orbit interaction results in the interconnection between electrical and spin currents~\cite{Dyakonov_Perel_1, Dyakonov_Perel_2}. An electric current $\mathbf{j} $ produces a transverse spin current $q_{ik} \sim \mathbf{j} $, leading to generation of spin density near the sample edges (the direct spin Hall effect; this term was coined out in Ref.~\cite{Hirsch}) and, vice versa, inhomogeneous distribution of spin density results in corrections to the electric current (the inverse spin Hall effect). In these phenomena, the relaxation of the electron spin and the spin current are of a large importance. The relaxation time of the first process is much more longer than the relaxation time of the second one~\cite{book}. The action of both the direct and the inverse spin Hall effects induces correction to the dc and the ac magnetoresistance in not too wide samples due to the formation of the near-edge layers with the perturbed spin density~$\mathbf{P} $~\cite{Dyakonov__spin_Hall_magnetores, Alekseev_Dyakonov__spin_Hall_magnetoimped}.

In pure metals with negligible density of defects, where the hydrodynamic regime of charge transport is realized, the effect of the rotational viscosity provide the interconnection between the electric current and spin density. Namely, the vorticity of the electron flow, $\nabla \times\,\mathbf{j}$, induces the difference in the non-diagonal components of the momentum flux density, $\Pi_{ik} \neq \Pi_{ki} $. Such difference leads to generation of the spin density~$\mathbf{P}$ due to conservation of the total angular momentum in electron-electron collisions~\cite{Maekawa_et_al__rotat_visc_1, Maekawa_et_al__rotat_visc_2, Maekawa_et_al__rotat_visc_3, Maekawa_et_al__rotat_visc_4, Polini__rotat_visc}. Similarly to spin Hall effects, rotational viscosity effect is related to the band spin-orbit interaction of conduction electrons. In particular, anomalous spin-orbit correction to the electron velocity leads to the mismatch of  $\Pi_{ik} $ and $ \Pi_{ki} $. In defectless samples, the spin current relaxation rate is of the order of the inter-particle scattering time $1/\tau_{ee}$, while the relaxation rate of the spin density is much lower, being typically proportional to $\Omega_{SO}^2\tau_{ee}$~\cite{Glazov_Ivchenko_1, Glazov_Ivchenko_2, D_Amico_Vignale____coulomb_drag_1, D_Amico_Vignale____coulomb_drag_2}, similarly to the case of Ohmic conductors (here $\Omega _{SO} $ is a small frequency of spin precession of conduction electrons due to the band spin-orbit interaction). Let us remind that in the Ohmic materials the relaxation time due to scattering on disorder, $\tau$, enters the precession spin relaxation rate, $\Omega_{SO}^2\tau$, instead of the the time $\tau_{ee}$ in defectless materials~\cite{Dyakonov_Perel_3__spin_rel}.

\begin{figure*}[t!]
	\centerline{\includegraphics[width=155mm]{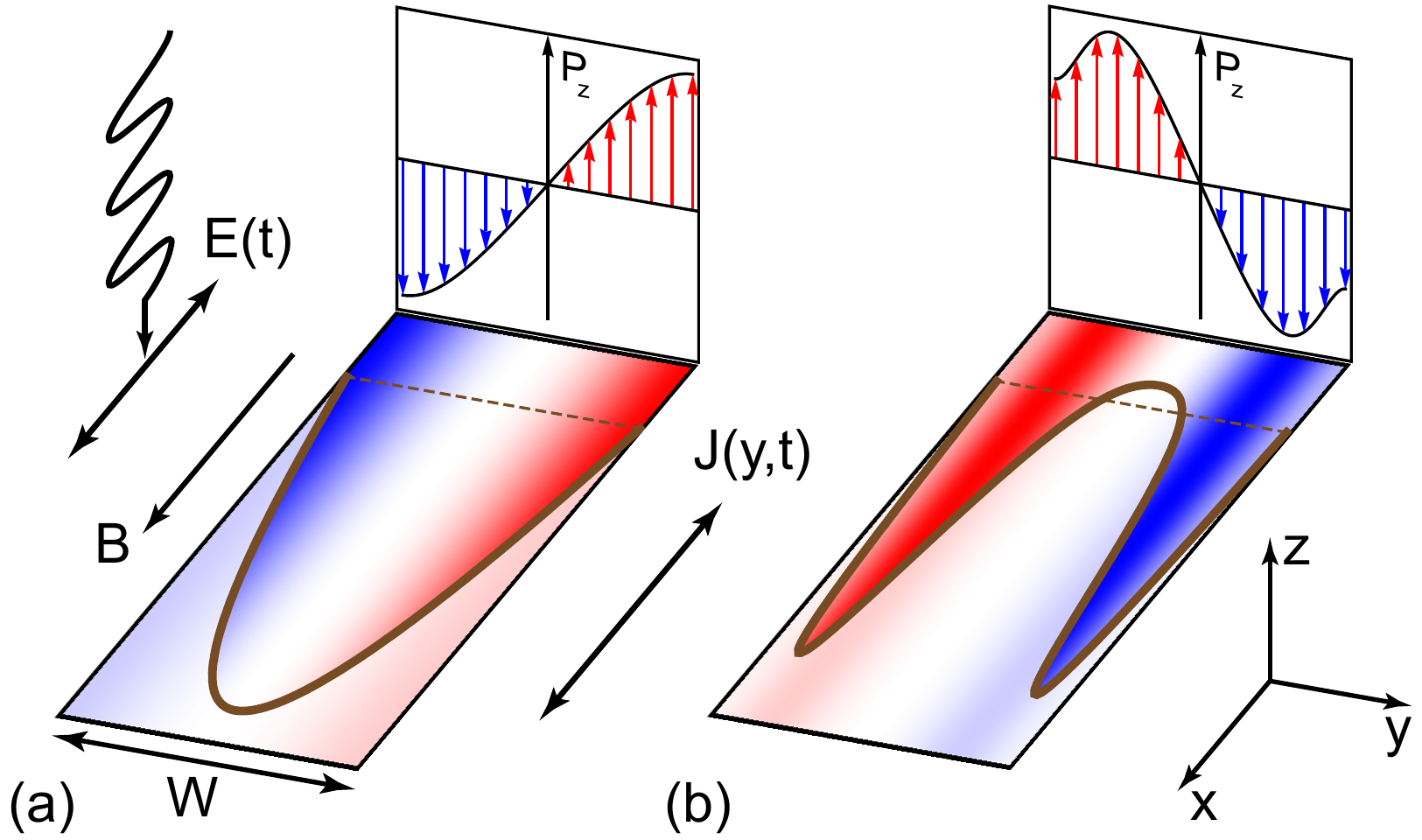}}
	\caption{\label{Fig_1} Dynamic generation of the spin polarization in a hydrodynamic flow of two-dimensional electrons in the presence of the lateral magnetic field and ac electric fields. The polarization density (red and blue) and current density (brown) profiles are shown for the quasi-static Poiseuille~\cite{Poiseuille}~(a) and  the strongly non-stationary Womersley~\cite{Womersley} regimes~(b).}
\end{figure*}

Recently, a thorough study of spin generation in hydrodynamic flows of the two-dimensional electron fluid was performed~\cite{Glazov___spin_Hall_in_hydr}. Various microscopic contribution to the interconnection between the spin current, the spin density and the electric current were calculated and compared. However, the relation between the rotational viscosity effect and microscopic mechanisms studied in Ref.~\cite{Glazov___spin_Hall_in_hydr} (for a particular geometry of the Poiseuille flow) remains unclear up to now. In this way, it is of interest to propose the experimental methods to determine the rotational viscosity coefficient and/or other kinetic coefficients responsible for the mutual transformation of the currents and the spin density in hydrodynamic conductors.

In this Letter we demonstrate that, under a high-frequency drive, a viscous flow of a two-dimensional electron fluid exhibits the electron spin resonance related to the own dynamics of the spin-polarized fluid in a weak magnetic field. The rotational viscosity effect leads to the dynamic spin-vorticity coupling, namely: the spin density generated by a viscous flow with a nonzero vorticity affects the flow itself resulting in a correction to the charge current density. In this way, we propose a purely electric method of observation of the spin resonance in hydrodynamic conductors, for example, in high-mobility GaAs quantum wells and graphene. The studied effect may allow to measure the rotational viscosity of the electron fluid in these materials.


{\em 2. High-frequency hydrodynamics of spin-polarized electron fluid.} We consider a flow of a two-dimensional electron fluid in a defectless sample [see Fig.~\ref{Fig_1}] in the external radio-frequency electric field $\mathbf{E}_0(t)$ and the magnetic field $\mathbf{B}$. We imply that the electrons form a Fermi gas. If the sample edges are rough and the scattering of electrons on them is diffusive, a non-stationary inhomogeneous viscous flow with the particle flow density $\mathbf{j}_ {\,0}(\mathbf{r},t) \sim \mathbf{E}_0$ is formed in the sample due to the shear viscosity effect~\cite{recentest___breathing_flow,Alekseev_Alekseeva}. Our aim is to calculate  the spin polarization of electrons $\mathbf{P}(\mathbf{r},t)$ in such flow, induced by the rotational viscosity, and to find the spin-related correction to the current density, $\mathbf{J}_2 (\mathbf{r},t) = e\,\mathbf{j}_2(\mathbf{r},t)$, and to the total electric current $\mathbf{I}_2(t) = \int _{\Sigma} dS_{\mathbf{r}} \, \mathbf{J}_2 (\mathbf{r},t) $ [here $e<0$ is the electron charge and $dS_{ \mathbf{r} } $ denotes differential along the one-dimensional section of the sample  $\Sigma$].

In such system the interconnection between $\mathbf{J} $ and $\mathbf{P} $ is controlled, in particular, by the rotational viscosity effect. This effect is microscopically induced by the band spin-orbit interaction of conduction electrons, leading to spin-dependent interparticle scattering and anomalous correction to the electron velocities~\cite{Maekawa_et_al__rotat_visc_1, Polini__rotat_visc,Glazov___spin_Hall_in_hydr}. This two processes establish the connection between the antisymmetric part of the momentum flux tensor $\Pi _{ik}$, $i \neq k$,  the curl of the flow, $\nabla \times \, \mathbf{j}$, which can be interpreted as the local rotational frequency $ \boldsymbol{\omega}_{orb} (\mathbf{r})$ of the fluid, and the spin polarization density $\mathbf{P}$~\cite{Maekawa_et_al__rotat_visc_1, Polini__rotat_visc}. The particular form of this connection is dictated by the symmetry of the system~\cite{DeGroot1962,Snider1967}. Rotational viscosity is the dissipative kinetic coefficient which interconnects the parts of the non-equilibrium thermodynamic flux ($\Pi_{ik}$) and forces ($\textbf{P}$ and $\partial_i j_k$) transforming under symmetry operations according to same one-dimensional irreducible representation~\cite{Snider1967}. For isotropic systems with $\mathrm{O}(3)$ symmetry this interconnection has the form~\cite{Polini__rotat_visc,DeGroot1962,Snider1967}:
\begin{equation}
	\label{Pi_a}
	\frac{\Pi_{ik}^a }{m}=
	\frac{\Pi_{ik} - \Pi_{ki}}{2 m}=
	\eta_r \epsilon _{ikl}\Big( [\nabla \times \mathbf{j}]  - \frac{\mu (\mathbf{P} - \mathbf{P} _0)}{\hbar} \Big)_l .
\end{equation}
For the two-dimensional fluid considered here we use the components of this equation describing the flows in $xy$-plane and three-dimensional spin polarization. Here $m$ is the electron mass, $\eta_r$ is the kinematic rotational viscosity coefficient, $ \epsilon _{ikl} $ is the unit antisymmetric tensor, $\mu$ is the total chemical potential of degenerate electrons, $\mathbf{P}_0 = \mathbf{S}_0 / n_0 = \chi \mathbf{B} /n_0$ is the equilibrium spin polarization, $\chi $ is the paramagnetic spin susceptibility of the electrons, and $n_0$ is the total electron density. In equation~(\ref{Pi_a}) the product $\mu P$ is the so-called spin chemical potential~\cite{Polini__rotat_visc} and, thus, $\mu P/ \hbar = (\mu_+ - \mu_-) / \hbar $ is the effective frequency of the spin rotation of electrons at the spin-splitted Fermi surfaces [here $\mu_\pm$ are their chemical potentials and the temperature is considered to be sufficiently low,~$T\ll \mu_+ - \mu_-$]. Microscopic calculations show~\cite{Polini__rotat_visc}, that in $\mathrm{A}_3 \mathrm{B}_5$ quantum wells the rotational viscosity coefficient $\eta_r$ is proportional to the squared Rashba coupling constant.

The hydrodynamic equations for the flow density $ \mathbf{j}$, the spin polarization $\mathbf{P} $, and the spin current $q_{ik}$ taking into account the shear viscosity and the rotational viscosity effects in a weak magnetic field can be written as
\begin{equation}
\label{main_system}
	\begin{array}{l}
	\displaystyle
	\frac{\partial \mathbf{j} }{\partial t}
	\, = \,
	\frac{ e \mathbf{E} _0(t) }{m} \, n_0
	\, + \,
	\eta \: \Delta \, \mathbf{j}
	\, - \,
	c_s^2 \, \frac{\eta_r }{ \eta_0 } \: [ \, \nabla \times\, \mathbf{P} \, ]
	\:,
	\\
	\\
	\displaystyle
	\frac{\partial P_k }{\partial t}
	\, + \,
	\frac{\partial q_{ik} }{\partial x_i}
	\, + \,
	[ \, \boldsymbol{\Omega} \times \mathbf{P} \, ]_k
	\, = \,
	\\	
	\\	
	\displaystyle
	\qquad\qquad
	\, = \, - \,
	\frac{ P_k - P_{0,k} }{ \tau_s }
	\, + \,
	2\frac{\eta_r }{ \eta_0 } \, [\, \nabla \times \, \mathbf{j} \, ]_k
	\:,
	\\
	\\
	\displaystyle
	\frac{\partial q_{ik} }{\partial t}
	\, = \, -
	c_s^2 \, \frac{\partial P_k}{\partial x_i}
	- \frac{ q_{ik} }{\tau_{sc}}
	\:,
	\end{array}
\end{equation}
where $\eta $ is the kinematic shear viscosity coefficient, $c_s = v_F/\sqrt{2}$ is the velocity of sound, $\eta_0 = \hbar / (2m) $ is the characteristic value of the dimension of viscosity, the ``quantum viscosity''~\cite{future2}, $\boldsymbol{\Omega} $ is the Larmor frequency due to the applied magnetic field $\mathbf{B}$, $\tau_s $ is the spin relaxation time related, for example, to the precession of the
electron spins in the spin-orbit effective magnetic field during the electron motion between electron-electron collisions~\cite{Glazov_Ivchenko_1, Glazov_Ivchenko_2} (for this mechanism $ 1/\tau_s \sim \Omega_{SO}^2\tau_{ee} \ll 1/\tau_{ee}$), $\tau_{sc} $ is the spin current relaxation time in electron-electron collisions due to spin Coulomb drag~\cite{D_Amico_Vignale____coulomb_drag_1, D_Amico_Vignale____coulomb_drag_2} (it was shown that $\tau_{sc} \sim \tau_{ee}$ in those works). The form of equations (\ref{main_system}) follows from consideration of Refs.~\cite{Maekawa_et_al__rotat_visc_1, Polini__rotat_visc}

The first line in Eqs.~(\ref{main_system}) is the hydrodynamic Navier-Stokes equation for the evolution of the mean electron momentum, containing the shear viscosity term and the rotational viscosity term. That is, the momentum flux tensor term, $- \partial \Pi_{il} /\partial x_l$, accounts for the two contributions: $\Pi_{ik} = \Pi_{ik}^s + \Pi_{ik}^a $, where $\Pi_{ik}^s = - m \, \eta \, (\partial j_i /\partial x_k + \partial j_k /\partial x_i)$ is the usual shear part of $\Pi_{ik}$ for incompressible flow, while $\Pi_{ik}^a$ is the rotational viscosity contribution given by Eq.~(\ref{Pi_a}).

The second of equations~(\ref{main_system}) describes the evolution of the electron spin, namely, its transfer by the spin current $q_{ik}$, rotation described by the Larmor frequency $\boldsymbol{\Omega}$, relaxation to the equilibrium value $\mathbf{S}_0 = n_0 \mathbf{P}_0 $, and the generation by the torque $T_k =\epsilon_{klq}\Pi^a_{lq}/m$ associated with the antisymmetric part of $\Pi_{ik}$~(\ref{Pi_a}). We imply that the relaxation term also contains the contribution from $\Pi_{ik}^a$

The third of equation~(\ref{main_system}) describes the evolution of the spin current $q_{ik}$. Provided that the spin current relaxation length, $v_F \tau_{sc} $, is sufficiently short, one can neglect the dissipative transfer of the spin flux in this equation (unlike the dissipative transfer of electron momentum described by $\Pi_{ik}$). Additionally, we omit possible contributions to the local variation of the spin current, $\partial q_{ik} /\partial t$, from the particle current $\mathbf{j}$, which are crucial for the spin Hall effect in Ohmic conductors~\cite{Dyakonov_Perel_1,Dyakonov_Perel_1,book}.

We consider slow flows with the frequencies $\omega \ll 1/\tau_{ee}$, thus the time dispersion of the viscosities $\eta$ and $\eta_r$ is insignificant. In this case, from the third of Eqs.~(\ref{main_system}) we obtain the direct connection between the spin polarization and the spin current: $q_{ik} = -D_s \, \partial P_k / \partial x_i$, where $D_s = c_s^2 \tau_{sc}$ is the spin diffusion coefficient.

The boundary condition describing the absence of the spin current at the closed part of the sample boundary, $\partial G $, takes the form $ (\, \partial P _k /\partial n \, ) |_{\,\partial G } = 0$, where $\mathbf{n}=\mathbf{n}(\mathbf{r})$ is the normal to the curve $\partial G $. The diffusive boundary condition at the closed part of the edges has the usual form: $ \mathbf{ j }|_{\,\partial G } = 0$.

\begin{figure}[t!]
	\centerline{\includegraphics[width=\linewidth]{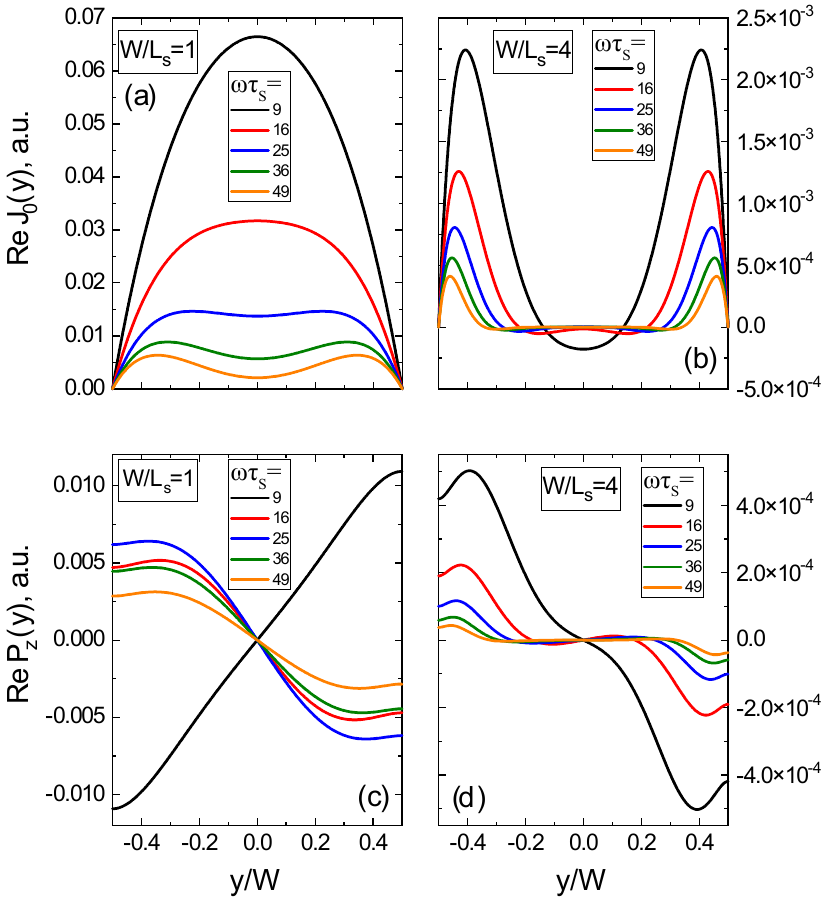}}
	\caption{\label{Fig_2}(a,b) Profiles of the amplitude $J_0(y)$ of the unperturbed electric current density $J_0(y,t) = J_0(y) e^{-i\omega t } +c.c.$ in a long sample in the parallel magnetic and electric field $\mathbf{E}_{0} (t)$ and $\mathbf{B}$ for  different parameters $W/L$ and $\omega \tau _s$.  Both values $J_0$ and $P_z$ are plotted in arbitrary units. (c,d) Profiles of the amplitude of the spin polarization $P_z (y,t) = P_z (y) e^{-i\omega t } +c.c.$ in the electron fluid, arising due to the rotational viscosity effect. The curves are plotted for the same sample at the same parameters as shown in panel~(a,b).}
\end{figure}


{\em 3. Spin resonance due to rotational viscosity.} First, we study the solutions of equations~(\ref{main_system}) in a long sample. The magnetic and the electric field both are directed along the sample: $\mathbf{E} _0(t) = \mathbf{E}_0 e^{-i\omega t} + c.c.$, $ \mathbf{E}_0 = E_0 \mathbf{e}_x$, $ \mathbf{B} = B \mathbf{e}_x$. In this geometry, all values are homogeneous along the $x$ coordinate and depend on the coordinate $y$ along the sample section: $P_y (y,t)$, $P_z (y,t)$, $ j= j_x (y,t) $ [see Fig.~\ref{Fig_1}]. The spin current tensor has the two non-zero components: $ q_{xy} $ and $ q_{xz} $.

The hydrodynamic equations (\ref{main_system}) take the form
\begin{equation}
	\label{hydr_eq_partic}
	\begin{array}{l}
	\displaystyle
	\Big( - i\omega - \eta \frac{d^2}{dy^2}\Big ) \, j \, = \, f_0
	\, + \, r_1 \, \frac{dP_z}{dy}
\:,
	\\
	\\
	\displaystyle
	\displaystyle
	\Big( - i\omega + \frac{1}{\tau_s} - D_s \frac{d^2}{dy^2}\Big ) \, P_\pm
	\, \pm \, i
	\Omega\, P _ \pm
	\, =\, r_2\, \frac{dj}{dy}
	\:,
	\end{array}
\end{equation}
where $P _ \pm = P_z \pm i P_y$, $f_0 = eE_0n_0 /m$, $r_1 = c_s^2\, \eta_r/ \eta_0 $, $r_2 = 2 \eta_r / \eta_0 $. The boundary conditions are formulated at the longitudinal edges: $ j|_{\, y = \pm W/2} =0 $ and $ (d P _ \pm /dy ) |_{\, y = \pm W/2} = 0 $.

The first line of Eqs.~(\ref{hydr_eq_partic}) with $\eta_r =0$ yields for the unperturbed particle flow: $j(y,t) = j(y) e^{-i\omega t } + c.c.$, where
\begin{equation}
	\label{q0}
	j (y) \, =\, f_0 \, \frac{i}{\omega} \, \Big[ 1-\frac{\cosh(\kappa y) }{\cosh(\kappa W/2)} \Big] \,, \quad \kappa = \sqrt { \, - i \, \frac{ \omega }{ \eta}} \,.
\end{equation}
Owing to the formula $\eta = v_F^2 \,  \tau_{2,ee}/4$, where $ \tau_{2,ee} $ is the shear stress relaxation time, we arrive to the estimate of the eigenvalue determining the characteristic width $L_\omega = 1/\mathrm{Re } \, \kappa $ of the near-edge layers of non-stationary Womersley flow~\cite{Womersley}: $\kappa \sim \sqrt{\omega / \tau_{ee} } \, /\,v_F$. In the static limit $\omega\to 0$, particle flow profile~(\ref{q0}) reduces to Poiseuille flow~\cite{Poiseuille}.

The eigenvalues of the unperturbed equations for the spin polarization [the second line of Eqs.~(\ref{hydr_eq_partic}) with $r_2 =0$] are
\begin{equation}
	\label{lambda_pm}
	\lambda_\pm \, = \, (1/L_s)\, \sqrt{\, 1 \, + \, i \, (\pm \Omega -\omega ) \, \tau_s \, } \:,
\end{equation}
where $L_s = \sqrt{D_s \tau_s} $ is the spin diffusion length. The relation between the shear viscosity and the spin characteristic lengths $L_\omega$ and $ L_ \lambda  = 1 / \mathrm{Re} \lambda _\pm $ is as follows:
 $L_\omega \sim L_ \lambda  $ far from the resonance frequency, $ |\pm \Omega -\omega | \, \tau_s \gg 1 $, and $L_\omega \ll L_ \lambda  $ near the resonance, $ |\pm \Omega -\omega | \, \tau_s \sim 1 $.

Circular components of the spin polarization generated by unperturbed flow~(\ref{q0}) are first order by $\eta_r$ and have the form
\begin{equation}
	\label{P}
	\begin{array}{c}
	\displaystyle
	P_\pm (y) =\frac{ i \, f_0 }{\omega} \, r_2\, \frac{ t_{\lambda _\pm } (y) -t_{ \kappa } ( y) }{\lambda _\pm ^2 - \kappa ^2 } \:,
	\end{array}
\end{equation}
where $t_{\xi} (y) = \sinh(\xi y) / [\xi \cosh(\xi W/2) ] $. The profiles of the electric current $J_0(y) =ej_0(y)$ and the $z$ component $P_z(y)$ of the spin polarization corresponding to Eqs.~(\ref{P}) are drawn in Fig.~\ref{Fig_2}.

Correction to the particle flow $j^{(2)}(y,t) = j^{(2)}(y) e^{-i\omega t} +c.c.$ generated by nonzero spin polarization~(\ref{P}) is second-order by $\eta_r$ and the equation for it is
\begin{equation}
	\label{eq_q2}
	\Big( - i\omega - \eta \frac{d^2}{dy^2}\Big ) \, j^{(2)} \, = \, r_1 \, \frac{dP_z}{dy} \:,
\end{equation}
where $ P_z = P_z^{(1)} $ is given by the sum of circular components $P_\pm $ after~(\ref{P}). The result of its solution is $j^{(2)} = (j^{(2)} _+ + j^{(2)} _- )/2 $, where
\begin{equation}
	\label{q_2}
	\begin{array}{c}
	\displaystyle j^{(2)}_\pm(y) \, = \,
	r_1 \, r_2 \, \frac{i f_0 }{ \omega } \frac{1}{\lambda_\pm^2 - \kappa^2 } \times
	\\
	\\
	\displaystyle
	\Big\{\, \frac{c_{\lambda_\pm}(y) - c_{\kappa}(y) }{\lambda_\pm^2 - \kappa^2 } - \frac{F_\kappa}{8} \Big[\, s_{\kappa}(y) \frac{y}{W/2}
- c_{\kappa}(y) \, \Big] \, \Big\}\, \:.
	\end{array}
\end{equation}
Here the notations $s_{\xi}(y) = \sinh(\xi y) / [\sinh(\xi W/2) ] $, $ c_{\xi}(y) = \cosh(\xi y) / [\cosh(\xi W/2) ] $, and $F_{\xi} = \tanh(\xi W/2 )/(\xi W/2)$ are used.

\begin{figure}[t!]
	\centerline{\includegraphics[width=\linewidth]{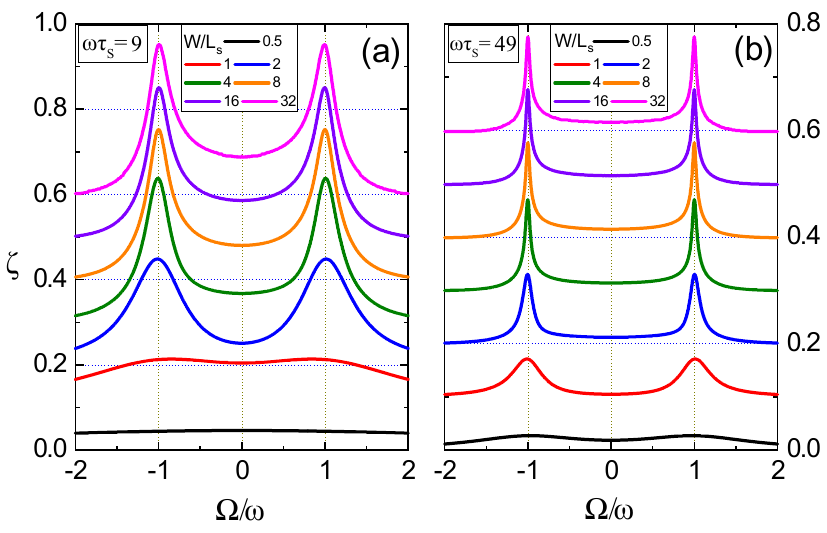}}
	\caption{\label{Fig_3} The relative spin-orbit correction to the real part of the impedance, $ \zeta = \mathrm{Re}\, \Delta Z (\Omega,\omega ) /\mathrm{Re}\, Z (\omega ) $, for the two-dimensional electron fluid in a long sample due to the rotational viscosity effect. This value is quadratic by the rotational viscosity coefficient: $\Delta Z \sim \eta_r^2$. The curves are plotted in arbitrary units for the different parameters $ W/L$ depicted on panels (a) for low and (b) high frequencies. For better visibility, the neighbor curves are shifted in the vertical direction by 0.1 at infinity, that leads to the almost same shifts at maximal values of $\Omega/\omega$, $\Omega/\omega = \pm 2$.}
\end{figure}

The main part of the total current $I (t) = e \int_{-W/2} ^{W/2} dy\, j (y,t)$ associated with~(\ref{q0}) is given by $I_0(t) = I_0 e^{-i\omega t} +c.c.$, where the amplitude is $ I_0 = (i \, e\, f_0 W / \omega) [ 1 - F(\kappa ) ]$. The spin-orbit correction $I^{(2)} (t)$ to it corresponding to $j^{(2)} (y)$~(\ref{q_2}) has the amplitude
\begin{equation}
	\label{I2}
	\begin{array}{c}
	\displaystyle
	\frac{ I^{(2)} }{W}  =  r_1  r_2   \frac{i e f_0  }{ \omega } \sum \limits_{\pm} \frac{1}{\lambda_\pm^2 - \kappa^2 }
 \Big(\, \frac{F_{\lambda_\pm} - F_\kappa }{\lambda_\pm^2 - \kappa^2 } - \frac{F'_\kappa}{2\kappa}    \, \Big) .
	\end{array}
\end{equation}
It is seen from Eqs.~(\ref{lambda_pm}) and (\ref{I2}) that $\mathrm{Re}\,I^{(2)} $ as a function of magnetic field exhibit the spin resonance for sufficiently narrow samples. This resonance originates from the formation of the near-edge layers with the perturbed spin polarization $\mathbf{P}(y,t)$, which induces the correction $I^{(2)} (y,t)$ to the current $I_0(t)$.

The real part of the correction to the mean sample impedance, $\Delta Z = E_0 W / (I_0 +I_2) - E_0W/I_0 $, is plotted in Fig.~\ref{Fig_3}. This value is proportional to the squared rotational viscosity, $r_1r_2 \sim \eta_r^2$ [see Eq.~(\ref{I2})]. We see from Fig.~\ref{Fig_3} that the higher is the ac frequency  and the wider is the sample, the narrower are the resonances at $\Omega = \pm \omega$. The limiting halfwidth at $W\gg L_s$ is estimated as $1/\tau_s$. The relative value of the correction to the impedance   $ \zeta = \mathrm{Re}\, \Delta Z (\Omega,\omega ) / \mathrm{Re}\, Z (\omega) = - \mathrm{Re}\, I_2 / \mathrm{Re}\, I_0 $ in sufficiently wide samples, $W\gg L_s$ is independent of the sample width. The frequency dependence of $ \zeta $ is determined by the values $\kappa(\omega)$ and $\lambda_\pm (\omega,\Omega) $ [see Eq.~(\ref{I2})].

The measurement of $\zeta (\Omega ) $ at fixed $\omega$, together with fitting of the whole curves $\zeta (\Omega)$ (see Fig.~\ref{Fig_3}), may allow to determine the rotational viscosity.


\begin{acknowledgements}

{\em 4. Conclusion and acknowledgments.} We have shown that the paramagnetic resonance of the conduction electrons can be observed in non-stationary viscous flow of two-dimensional electron fluid owing to the rotational viscosity effect. The resonance manifest itself in the spin-orbit correction to the sample impedance. Measurements of this resonance may allow to determine the rotational viscosity coefficient of the viscous electron fluid.

We thank M.~I.~Dyakonov for drawing our attention to the problem studied in this work and for fruitful  discussions of some of the issues raised in this work. We thank M.~M.~Glazov for  informing us about the content of work~\cite{Glazov___spin_Hall_in_hydr} prior to its publication and fruitful  discussions. This study was supported by the Russian Foundation for Basic Research (Grant No. 19-02-00999) and by the Basis Foundation (Grant No. 19-1-5-127-1 and 20-1-3-51-1).

\end{acknowledgements}

\bibliography{RVSR_bib}

\begin{thebibliography}{65}%
\makeatletter
\providecommand \@ifxundefined [1]{%
 \@ifx{#1\undefined}
}%
\providecommand \@ifnum [1]{%
 \ifnum #1\expandafter \@firstoftwo
 \else \expandafter \@secondoftwo
 \fi
}%
\providecommand \@ifx [1]{%
 \ifx #1\expandafter \@firstoftwo
 \else \expandafter \@secondoftwo
 \fi
}%
\providecommand \natexlab [1]{#1}%
\providecommand \enquote  [1]{``#1''}%
\providecommand \bibnamefont  [1]{#1}%
\providecommand \bibfnamefont [1]{#1}%
\providecommand \citenamefont [1]{#1}%
\providecommand \href@noop [0]{\@secondoftwo}%
\providecommand \href [0]{\begingroup \@sanitize@url \@href}%
\providecommand \@href[1]{\@@startlink{#1}\@@href}%
\providecommand \@@href[1]{\endgroup#1\@@endlink}%
\providecommand \@sanitize@url [0]{\catcode `\\12\catcode `\$12\catcode
  `\&12\catcode `\#12\catcode `\^12\catcode `\_12\catcode `\%12\relax}%
\providecommand \@@startlink[1]{}%
\providecommand \@@endlink[0]{}%
\providecommand \url  [0]{\begingroup\@sanitize@url \@url }%
\providecommand \@url [1]{\endgroup\@href {#1}{\urlprefix }}%
\providecommand \urlprefix  [0]{URL }%
\providecommand \Eprint [0]{\href }%
\providecommand \doibase [0]{https://doi.org/}%
\providecommand \selectlanguage [0]{\@gobble}%
\providecommand \bibinfo  [0]{\@secondoftwo}%
\providecommand \bibfield  [0]{\@secondoftwo}%
\providecommand \translation [1]{[#1]}%
\providecommand \BibitemOpen [0]{}%
\providecommand \bibitemStop [0]{}%
\providecommand \bibitemNoStop [0]{.\EOS\space}%
\providecommand \EOS [0]{\spacefactor3000\relax}%
\providecommand \BibitemShut  [1]{\csname bibitem#1\endcsname}%
\let\auto@bib@innerbib\@empty
\bibitem [{\citenamefont {Gurzhi}(1968)}]{Gurzhi}%
  \BibitemOpen
  \bibfield  {author} {\bibinfo {author} {\bibfnamefont {R.~N.}\ \bibnamefont
  {Gurzhi}},\ }\href {https://doi.org/10.1070/PU1968v011n02ABEH003815}
  {\bibfield  {journal} {\bibinfo  {journal} {Sov. Phys. Uspekhi}\ }\textbf
  {\bibinfo {volume} {11}},\ \bibinfo {pages} {255} (\bibinfo {year}
  {1968})}\BibitemShut {NoStop}%
\bibitem [{\citenamefont {Bandurin}\ \emph {et~al.}(2016)\citenamefont
  {Bandurin}, \citenamefont {Torre}, \citenamefont {Kumar}, \citenamefont
  {Ben~Shalom}, \citenamefont {Tomadin}, \citenamefont {Principi},
  \citenamefont {Auton}, \citenamefont {Khestanova}, \citenamefont {Novoselov},
  \citenamefont {Grigorieva}, \citenamefont {Ponomarenko}, \citenamefont
  {Geim},\ and\ \citenamefont {Polini}}]{grahene}%
  \BibitemOpen
  \bibfield  {author} {\bibinfo {author} {\bibfnamefont {D.~A.}\ \bibnamefont
  {Bandurin}}, \bibinfo {author} {\bibfnamefont {I.}~\bibnamefont {Torre}},
  \bibinfo {author} {\bibfnamefont {R.~K.}\ \bibnamefont {Kumar}}, \bibinfo
  {author} {\bibfnamefont {M.}~\bibnamefont {Ben~Shalom}}, \bibinfo {author}
  {\bibfnamefont {A.}~\bibnamefont {Tomadin}}, \bibinfo {author} {\bibfnamefont
  {A.}~\bibnamefont {Principi}}, \bibinfo {author} {\bibfnamefont {G.~H.}\
  \bibnamefont {Auton}}, \bibinfo {author} {\bibfnamefont {E.}~\bibnamefont
  {Khestanova}}, \bibinfo {author} {\bibfnamefont {K.~S.}\ \bibnamefont
  {Novoselov}}, \bibinfo {author} {\bibfnamefont {I.~V.}\ \bibnamefont
  {Grigorieva}}, \bibinfo {author} {\bibfnamefont {L.~A.}\ \bibnamefont
  {Ponomarenko}}, \bibinfo {author} {\bibfnamefont {A.~K.}\ \bibnamefont
  {Geim}},\ and\ \bibinfo {author} {\bibfnamefont {M.}~\bibnamefont {Polini}},\
  }\href {https://doi.org/10.1126/science.aad0201} {\bibfield  {journal}
  {\bibinfo  {journal} {Science}\ }\textbf {\bibinfo {volume} {351}},\ \bibinfo
  {pages} {1055} (\bibinfo {year} {2016})}\BibitemShut {NoStop}%
\bibitem [{\citenamefont {Krishna~Kumar}\ \emph {et~al.}(2017)\citenamefont
  {Krishna~Kumar}, \citenamefont {Bandurin}, \citenamefont {Pellegrino},
  \citenamefont {Cao}, \citenamefont {Principi}, \citenamefont {Guo},
  \citenamefont {Auton}, \citenamefont {Ben~Shalom}, \citenamefont
  {Ponomarenko}, \citenamefont {Falkovich}, \citenamefont {Watanabe},
  \citenamefont {Taniguchi}, \citenamefont {Grigorieva}, \citenamefont
  {Levitov}, \citenamefont {Polini},\ and\ \citenamefont {Geim}}]{grahene_2}%
  \BibitemOpen
  \bibfield  {author} {\bibinfo {author} {\bibfnamefont {R.}~\bibnamefont
  {Krishna~Kumar}}, \bibinfo {author} {\bibfnamefont {D.~A.}\ \bibnamefont
  {Bandurin}}, \bibinfo {author} {\bibfnamefont {F.~M.~D.}\ \bibnamefont
  {Pellegrino}}, \bibinfo {author} {\bibfnamefont {Y.}~\bibnamefont {Cao}},
  \bibinfo {author} {\bibfnamefont {A.}~\bibnamefont {Principi}}, \bibinfo
  {author} {\bibfnamefont {H.}~\bibnamefont {Guo}}, \bibinfo {author}
  {\bibfnamefont {G.~H.}\ \bibnamefont {Auton}}, \bibinfo {author}
  {\bibfnamefont {M.}~\bibnamefont {Ben~Shalom}}, \bibinfo {author}
  {\bibfnamefont {L.~A.}\ \bibnamefont {Ponomarenko}}, \bibinfo {author}
  {\bibfnamefont {G.}~\bibnamefont {Falkovich}}, \bibinfo {author}
  {\bibfnamefont {K.}~\bibnamefont {Watanabe}}, \bibinfo {author}
  {\bibfnamefont {T.}~\bibnamefont {Taniguchi}}, \bibinfo {author}
  {\bibfnamefont {I.~V.}\ \bibnamefont {Grigorieva}}, \bibinfo {author}
  {\bibfnamefont {L.~S.}\ \bibnamefont {Levitov}}, \bibinfo {author}
  {\bibfnamefont {M.}~\bibnamefont {Polini}},\ and\ \bibinfo {author}
  {\bibfnamefont {A.~K.}\ \bibnamefont {Geim}},\ }\href
  {https://doi.org/10.1038/nphys4240} {\bibfield  {journal} {\bibinfo
  {journal} {Nature Physics}\ }\textbf {\bibinfo {volume} {13}},\ \bibinfo
  {pages} {1182} (\bibinfo {year} {2017})}\BibitemShut {NoStop}%
\bibitem [{\citenamefont {Berdyugin}\ \emph {et~al.}(2019)\citenamefont
  {Berdyugin}, \citenamefont {Xu}, \citenamefont {Pellegrino}, \citenamefont
  {Krishna~Kumar}, \citenamefont {Principi}, \citenamefont {Torre},
  \citenamefont {Ben~Shalom}, \citenamefont {Taniguchi}, \citenamefont
  {Watanabe}, \citenamefont {Grigorieva}, \citenamefont {Polini}, \citenamefont
  {Geim},\ and\ \citenamefont {Bandurin}}]{grahene_3}%
  \BibitemOpen
  \bibfield  {author} {\bibinfo {author} {\bibfnamefont {A.~I.}\ \bibnamefont
  {Berdyugin}}, \bibinfo {author} {\bibfnamefont {S.~G.}\ \bibnamefont {Xu}},
  \bibinfo {author} {\bibfnamefont {F.~M.~D.}\ \bibnamefont {Pellegrino}},
  \bibinfo {author} {\bibfnamefont {R.}~\bibnamefont {Krishna~Kumar}}, \bibinfo
  {author} {\bibfnamefont {A.}~\bibnamefont {Principi}}, \bibinfo {author}
  {\bibfnamefont {I.}~\bibnamefont {Torre}}, \bibinfo {author} {\bibfnamefont
  {M.}~\bibnamefont {Ben~Shalom}}, \bibinfo {author} {\bibfnamefont
  {T.}~\bibnamefont {Taniguchi}}, \bibinfo {author} {\bibfnamefont
  {K.}~\bibnamefont {Watanabe}}, \bibinfo {author} {\bibfnamefont {I.~V.}\
  \bibnamefont {Grigorieva}}, \bibinfo {author} {\bibfnamefont
  {M.}~\bibnamefont {Polini}}, \bibinfo {author} {\bibfnamefont {A.~K.}\
  \bibnamefont {Geim}},\ and\ \bibinfo {author} {\bibfnamefont {D.~A.}\
  \bibnamefont {Bandurin}},\ }\href {https://doi.org/10.1126/science.aau0685}
  {\bibfield  {journal} {\bibinfo  {journal} {Science}\ }\textbf {\bibinfo
  {volume} {364}},\ \bibinfo {pages} {162} (\bibinfo {year}
  {2019})}\BibitemShut {NoStop}%
\bibitem [{\citenamefont {Levitov}\ and\ \citenamefont
  {Falkovich}(2016)}]{Levitov_et_al}%
  \BibitemOpen
  \bibfield  {author} {\bibinfo {author} {\bibfnamefont {L.}~\bibnamefont
  {Levitov}}\ and\ \bibinfo {author} {\bibfnamefont {G.}~\bibnamefont
  {Falkovich}},\ }\href {https://doi.org/10.1038/nphys3667} {\bibfield
  {journal} {\bibinfo  {journal} {Nature Physics}\ }\textbf {\bibinfo {volume}
  {12}},\ \bibinfo {pages} {672} (\bibinfo {year} {2016})}\BibitemShut
  {NoStop}%
\bibitem [{\citenamefont {Sulpizio}\ \emph {et~al.}(2019)\citenamefont
  {Sulpizio}, \citenamefont {Ella}, \citenamefont {Rozen}, \citenamefont
  {Birkbeck}, \citenamefont {Perello}, \citenamefont {Dutta}, \citenamefont
  {Ben-Shalom}, \citenamefont {Taniguchi}, \citenamefont {Watanabe},
  \citenamefont {Holder}, \citenamefont {Queiroz}, \citenamefont {Principi},
  \citenamefont {Stern}, \citenamefont {Scaffidi}, \citenamefont {Geim},\ and\
  \citenamefont {Ilani}}]{profile_1}%
  \BibitemOpen
  \bibfield  {author} {\bibinfo {author} {\bibfnamefont {J.~A.}\ \bibnamefont
  {Sulpizio}}, \bibinfo {author} {\bibfnamefont {L.}~\bibnamefont {Ella}},
  \bibinfo {author} {\bibfnamefont {A.}~\bibnamefont {Rozen}}, \bibinfo
  {author} {\bibfnamefont {J.}~\bibnamefont {Birkbeck}}, \bibinfo {author}
  {\bibfnamefont {D.~J.}\ \bibnamefont {Perello}}, \bibinfo {author}
  {\bibfnamefont {D.}~\bibnamefont {Dutta}}, \bibinfo {author} {\bibfnamefont
  {M.}~\bibnamefont {Ben-Shalom}}, \bibinfo {author} {\bibfnamefont
  {T.}~\bibnamefont {Taniguchi}}, \bibinfo {author} {\bibfnamefont
  {K.}~\bibnamefont {Watanabe}}, \bibinfo {author} {\bibfnamefont
  {T.}~\bibnamefont {Holder}}, \bibinfo {author} {\bibfnamefont
  {R.}~\bibnamefont {Queiroz}}, \bibinfo {author} {\bibfnamefont
  {A.}~\bibnamefont {Principi}}, \bibinfo {author} {\bibfnamefont
  {A.}~\bibnamefont {Stern}}, \bibinfo {author} {\bibfnamefont
  {T.}~\bibnamefont {Scaffidi}}, \bibinfo {author} {\bibfnamefont {A.~K.}\
  \bibnamefont {Geim}},\ and\ \bibinfo {author} {\bibfnamefont
  {S.}~\bibnamefont {Ilani}},\ }\href
  {https://doi.org/10.1038/s41586-019-1788-9} {\bibfield  {journal} {\bibinfo
  {journal} {Nature}\ }\textbf {\bibinfo {volume} {576}},\ \bibinfo {pages}
  {75} (\bibinfo {year} {2019})}\BibitemShut {NoStop}%
\bibitem [{\citenamefont {Ku}\ \emph {et~al.}(2019)\citenamefont {Ku},
  \citenamefont {Zhou}, \citenamefont {Li}, \citenamefont {Shin}, \citenamefont
  {Shi}, \citenamefont {Burch}, \citenamefont {Anderson}, \citenamefont
  {Pierce}, \citenamefont {Xie}, \citenamefont {Hamo}, \citenamefont {Vool},
  \citenamefont {Zhang}, \citenamefont {Casola}, \citenamefont {Taniguchi},
  \citenamefont {Watanabe}, \citenamefont {Fogler}, \citenamefont {Kim},
  \citenamefont {Yacoby},\ and\ \citenamefont {Walsworth}}]{profile_2}%
  \BibitemOpen
  \bibfield  {author} {\bibinfo {author} {\bibfnamefont {M.~J.~H.}\
  \bibnamefont {Ku}}, \bibinfo {author} {\bibfnamefont {T.~X.}\ \bibnamefont
  {Zhou}}, \bibinfo {author} {\bibfnamefont {Q.}~\bibnamefont {Li}}, \bibinfo
  {author} {\bibfnamefont {Y.~J.}\ \bibnamefont {Shin}}, \bibinfo {author}
  {\bibfnamefont {J.~K.}\ \bibnamefont {Shi}}, \bibinfo {author} {\bibfnamefont
  {C.}~\bibnamefont {Burch}}, \bibinfo {author} {\bibfnamefont {L.~E.}\
  \bibnamefont {Anderson}}, \bibinfo {author} {\bibfnamefont {A.~T.}\
  \bibnamefont {Pierce}}, \bibinfo {author} {\bibfnamefont {Y.}~\bibnamefont
  {Xie}}, \bibinfo {author} {\bibfnamefont {A.}~\bibnamefont {Hamo}}, \bibinfo
  {author} {\bibfnamefont {U.}~\bibnamefont {Vool}}, \bibinfo {author}
  {\bibfnamefont {H.}~\bibnamefont {Zhang}}, \bibinfo {author} {\bibfnamefont
  {F.}~\bibnamefont {Casola}}, \bibinfo {author} {\bibfnamefont
  {T.}~\bibnamefont {Taniguchi}}, \bibinfo {author} {\bibfnamefont
  {K.}~\bibnamefont {Watanabe}}, \bibinfo {author} {\bibfnamefont {M.~M.}\
  \bibnamefont {Fogler}}, \bibinfo {author} {\bibfnamefont {P.}~\bibnamefont
  {Kim}}, \bibinfo {author} {\bibfnamefont {A.}~\bibnamefont {Yacoby}},\ and\
  \bibinfo {author} {\bibfnamefont {R.~L.}\ \bibnamefont {Walsworth}},\ }\href
  {https://doi.org/10.1038/s41586-020-2507-2} {\bibfield  {journal} {\bibinfo
  {journal} {Nature}\ }\textbf {\bibinfo {volume} {583}},\ \bibinfo {pages}
  {537} (\bibinfo {year} {2019})}\BibitemShut {NoStop}%
\bibitem [{\citenamefont {Moll}\ \emph {et~al.}(2016)\citenamefont {Moll},
  \citenamefont {Kushwaha}, \citenamefont {Nandi}, \citenamefont {Schmidt},\
  and\ \citenamefont {Mackenzie}}]{Weyl_sem_1}%
  \BibitemOpen
  \bibfield  {author} {\bibinfo {author} {\bibfnamefont {P.~J.~W.}\
  \bibnamefont {Moll}}, \bibinfo {author} {\bibfnamefont {P.}~\bibnamefont
  {Kushwaha}}, \bibinfo {author} {\bibfnamefont {N.}~\bibnamefont {Nandi}},
  \bibinfo {author} {\bibfnamefont {B.}~\bibnamefont {Schmidt}},\ and\ \bibinfo
  {author} {\bibfnamefont {A.~P.}\ \bibnamefont {Mackenzie}},\ }\href
  {https://doi.org/10.1126/science.aac8385} {\bibfield  {journal} {\bibinfo
  {journal} {Science}\ }\textbf {\bibinfo {volume} {351}},\ \bibinfo {pages}
  {1061} (\bibinfo {year} {2016})}\BibitemShut {NoStop}%
\bibitem [{\citenamefont {Gooth}\ \emph {et~al.}(2018)\citenamefont {Gooth},
  \citenamefont {Menges}, \citenamefont {Kumar}, \citenamefont {S{\"u}{\ss}},
  \citenamefont {Shekhar}, \citenamefont {Sun}, \citenamefont {Drechsler},
  \citenamefont {Zierold}, \citenamefont {Felser},\ and\ \citenamefont
  {Gotsmann}}]{Weyl_sem_2}%
  \BibitemOpen
  \bibfield  {author} {\bibinfo {author} {\bibfnamefont {J.}~\bibnamefont
  {Gooth}}, \bibinfo {author} {\bibfnamefont {F.}~\bibnamefont {Menges}},
  \bibinfo {author} {\bibfnamefont {N.}~\bibnamefont {Kumar}}, \bibinfo
  {author} {\bibfnamefont {V.}~\bibnamefont {S{\"u}{\ss}}}, \bibinfo {author}
  {\bibfnamefont {C.}~\bibnamefont {Shekhar}}, \bibinfo {author} {\bibfnamefont
  {Y.}~\bibnamefont {Sun}}, \bibinfo {author} {\bibfnamefont {U.}~\bibnamefont
  {Drechsler}}, \bibinfo {author} {\bibfnamefont {R.}~\bibnamefont {Zierold}},
  \bibinfo {author} {\bibfnamefont {C.}~\bibnamefont {Felser}},\ and\ \bibinfo
  {author} {\bibfnamefont {B.}~\bibnamefont {Gotsmann}},\ }\href
  {https://doi.org/10.1038/s41467-018-06688-y} {\bibfield  {journal} {\bibinfo
  {journal} {Nature Communications}\ }\textbf {\bibinfo {volume} {9}},\
  \bibinfo {pages} {4093} (\bibinfo {year} {2018})}\BibitemShut {NoStop}%
\bibitem [{\citenamefont {Hatke}\ \emph {et~al.}(2012)\citenamefont {Hatke},
  \citenamefont {Zudov}, \citenamefont {Reno}, \citenamefont {Pfeiffer},\ and\
  \citenamefont {West}}]{exps_neg_1}%
  \BibitemOpen
  \bibfield  {author} {\bibinfo {author} {\bibfnamefont {A.~T.}\ \bibnamefont
  {Hatke}}, \bibinfo {author} {\bibfnamefont {M.~A.}\ \bibnamefont {Zudov}},
  \bibinfo {author} {\bibfnamefont {J.~L.}\ \bibnamefont {Reno}}, \bibinfo
  {author} {\bibfnamefont {L.~N.}\ \bibnamefont {Pfeiffer}},\ and\ \bibinfo
  {author} {\bibfnamefont {K.~W.}\ \bibnamefont {West}},\ }\href
  {https://doi.org/10.1103/PhysRevB.85.081304} {\bibfield  {journal} {\bibinfo
  {journal} {Phys. Rev. B}\ }\textbf {\bibinfo {volume} {85}},\ \bibinfo
  {pages} {081304(R)} (\bibinfo {year} {2012})}\BibitemShut {NoStop}%
\bibitem [{\citenamefont {Mani}\ \emph {et~al.}(2013)\citenamefont {Mani},
  \citenamefont {Kriisa},\ and\ \citenamefont {Wegscheider}}]{exps_neg_2}%
  \BibitemOpen
  \bibfield  {author} {\bibinfo {author} {\bibfnamefont {R.~G.}\ \bibnamefont
  {Mani}}, \bibinfo {author} {\bibfnamefont {A.}~\bibnamefont {Kriisa}},\ and\
  \bibinfo {author} {\bibfnamefont {W.}~\bibnamefont {Wegscheider}},\ }\href
  {https://doi.org/10.1038/srep02747} {\bibfield  {journal} {\bibinfo
  {journal} {Scientific Reports}\ }\textbf {\bibinfo {volume} {3}},\ \bibinfo
  {pages} {2747} (\bibinfo {year} {2013})}\BibitemShut {NoStop}%
\bibitem [{\citenamefont {Bockhorn}\ \emph {et~al.}(2011)\citenamefont
  {Bockhorn}, \citenamefont {Barthold}, \citenamefont {Schuh}, \citenamefont
  {Wegscheider},\ and\ \citenamefont {Haug}}]{exps_neg_3}%
  \BibitemOpen
  \bibfield  {author} {\bibinfo {author} {\bibfnamefont {L.}~\bibnamefont
  {Bockhorn}}, \bibinfo {author} {\bibfnamefont {P.}~\bibnamefont {Barthold}},
  \bibinfo {author} {\bibfnamefont {D.}~\bibnamefont {Schuh}}, \bibinfo
  {author} {\bibfnamefont {W.}~\bibnamefont {Wegscheider}},\ and\ \bibinfo
  {author} {\bibfnamefont {R.~J.}\ \bibnamefont {Haug}},\ }\href
  {https://doi.org/10.1103/PhysRevB.83.113301} {\bibfield  {journal} {\bibinfo
  {journal} {Phys. Rev. B}\ }\textbf {\bibinfo {volume} {83}},\ \bibinfo
  {pages} {113301} (\bibinfo {year} {2011})}\BibitemShut {NoStop}%
\bibitem [{\citenamefont {Shi}\ \emph {et~al.}(2014)\citenamefont {Shi},
  \citenamefont {Martin}, \citenamefont {Ebner}, \citenamefont {Zudov},
  \citenamefont {Pfeiffer},\ and\ \citenamefont {West}}]{exps_neg_4}%
  \BibitemOpen
  \bibfield  {author} {\bibinfo {author} {\bibfnamefont {Q.}~\bibnamefont
  {Shi}}, \bibinfo {author} {\bibfnamefont {P.~D.}\ \bibnamefont {Martin}},
  \bibinfo {author} {\bibfnamefont {Q.~A.}\ \bibnamefont {Ebner}}, \bibinfo
  {author} {\bibfnamefont {M.~A.}\ \bibnamefont {Zudov}}, \bibinfo {author}
  {\bibfnamefont {L.~N.}\ \bibnamefont {Pfeiffer}},\ and\ \bibinfo {author}
  {\bibfnamefont {K.~W.}\ \bibnamefont {West}},\ }\href
  {https://doi.org/10.1103/PhysRevB.89.201301} {\bibfield  {journal} {\bibinfo
  {journal} {Phys. Rev. B}\ }\textbf {\bibinfo {volume} {89}},\ \bibinfo
  {pages} {201301(R)} (\bibinfo {year} {2014})}\BibitemShut {NoStop}%
\bibitem [{\citenamefont {Alekseev}(2016)}]{je_visc}%
  \BibitemOpen
  \bibfield  {author} {\bibinfo {author} {\bibfnamefont {P.~S.}\ \bibnamefont
  {Alekseev}},\ }\href {https://doi.org/10.1103/PhysRevLett.117.166601}
  {\bibfield  {journal} {\bibinfo  {journal} {Phys. Rev. Lett.}\ }\textbf
  {\bibinfo {volume} {117}},\ \bibinfo {pages} {166601} (\bibinfo {year}
  {2016})}\BibitemShut {NoStop}%
\bibitem [{\citenamefont {Gusev}\ \emph
  {et~al.}(2018{\natexlab{a}})\citenamefont {Gusev}, \citenamefont {Levin},
  \citenamefont {Levinson},\ and\ \citenamefont {Bakarov}}]{Gusev_1}%
  \BibitemOpen
  \bibfield  {author} {\bibinfo {author} {\bibfnamefont {G.~M.}\ \bibnamefont
  {Gusev}}, \bibinfo {author} {\bibfnamefont {A.~D.}\ \bibnamefont {Levin}},
  \bibinfo {author} {\bibfnamefont {E.~V.}\ \bibnamefont {Levinson}},\ and\
  \bibinfo {author} {\bibfnamefont {A.~K.}\ \bibnamefont {Bakarov}},\ }\href
  {https://doi.org/10.1063/1.5020763} {\bibfield  {journal} {\bibinfo
  {journal} {AIP Advances}\ }\textbf {\bibinfo {volume} {8}},\ \bibinfo {pages}
  {025318} (\bibinfo {year} {2018}{\natexlab{a}})}\BibitemShut {NoStop}%
\bibitem [{\citenamefont {Levin}\ \emph {et~al.}(2018)\citenamefont {Levin},
  \citenamefont {Gusev}, \citenamefont {Levinson}, \citenamefont {Kvon},\ and\
  \citenamefont {Bakarov}}]{Gusev_2}%
  \BibitemOpen
  \bibfield  {author} {\bibinfo {author} {\bibfnamefont {A.~D.}\ \bibnamefont
  {Levin}}, \bibinfo {author} {\bibfnamefont {G.~M.}\ \bibnamefont {Gusev}},
  \bibinfo {author} {\bibfnamefont {E.~V.}\ \bibnamefont {Levinson}}, \bibinfo
  {author} {\bibfnamefont {Z.~D.}\ \bibnamefont {Kvon}},\ and\ \bibinfo
  {author} {\bibfnamefont {A.~K.}\ \bibnamefont {Bakarov}},\ }\href
  {https://doi.org/10.1103/PhysRevB.97.245308} {\bibfield  {journal} {\bibinfo
  {journal} {Phys. Rev. B}\ }\textbf {\bibinfo {volume} {97}},\ \bibinfo
  {pages} {245308} (\bibinfo {year} {2018})}\BibitemShut {NoStop}%
\bibitem [{\citenamefont {Gusev}\ \emph
  {et~al.}(2018{\natexlab{b}})\citenamefont {Gusev}, \citenamefont {Levin},
  \citenamefont {Levinson},\ and\ \citenamefont {Bakarov}}]{Gusev_3}%
  \BibitemOpen
  \bibfield  {author} {\bibinfo {author} {\bibfnamefont {G.~M.}\ \bibnamefont
  {Gusev}}, \bibinfo {author} {\bibfnamefont {A.~D.}\ \bibnamefont {Levin}},
  \bibinfo {author} {\bibfnamefont {E.~V.}\ \bibnamefont {Levinson}},\ and\
  \bibinfo {author} {\bibfnamefont {A.~K.}\ \bibnamefont {Bakarov}},\ }\href
  {https://doi.org/10.1103/PhysRevB.98.161303} {\bibfield  {journal} {\bibinfo
  {journal} {Phys. Rev. B}\ }\textbf {\bibinfo {volume} {98}},\ \bibinfo
  {pages} {161303(R)} (\bibinfo {year} {2018}{\natexlab{b}})}\BibitemShut
  {NoStop}%
\bibitem [{\citenamefont {Keser}\ \emph {et~al.}(2021)\citenamefont {Keser},
  \citenamefont {Wang}, \citenamefont {Klochan}, \citenamefont {Ho},
  \citenamefont {Tkachenko}, \citenamefont {Tkachenko}, \citenamefont {Culcer},
  \citenamefont {Adam}, \citenamefont {Farrer}, \citenamefont {Ritchie},
  \citenamefont {Sushkov},\ and\ \citenamefont {Hamilton}}]{recent___1}%
  \BibitemOpen
  \bibfield  {author} {\bibinfo {author} {\bibfnamefont {A.~C.}\ \bibnamefont
  {Keser}}, \bibinfo {author} {\bibfnamefont {D.~Q.}\ \bibnamefont {Wang}},
  \bibinfo {author} {\bibfnamefont {O.}~\bibnamefont {Klochan}}, \bibinfo
  {author} {\bibfnamefont {D.~Y.~H.}\ \bibnamefont {Ho}}, \bibinfo {author}
  {\bibfnamefont {O.~A.}\ \bibnamefont {Tkachenko}}, \bibinfo {author}
  {\bibfnamefont {V.~A.}\ \bibnamefont {Tkachenko}}, \bibinfo {author}
  {\bibfnamefont {D.}~\bibnamefont {Culcer}}, \bibinfo {author} {\bibfnamefont
  {S.}~\bibnamefont {Adam}}, \bibinfo {author} {\bibfnamefont {I.}~\bibnamefont
  {Farrer}}, \bibinfo {author} {\bibfnamefont {D.~A.}\ \bibnamefont {Ritchie}},
  \bibinfo {author} {\bibfnamefont {O.~P.}\ \bibnamefont {Sushkov}},\ and\
  \bibinfo {author} {\bibfnamefont {A.~R.}\ \bibnamefont {Hamilton}},\ }\href
  {https://doi.org/10.1103/PhysRevX.11.031030} {\bibfield  {journal} {\bibinfo
  {journal} {Phys. Rev. X}\ }\textbf {\bibinfo {volume} {11}},\ \bibinfo
  {pages} {031030} (\bibinfo {year} {2021})}\BibitemShut {NoStop}%
\bibitem [{\citenamefont {Gupta}\ \emph {et~al.}(2021)\citenamefont {Gupta},
  \citenamefont {Heremans}, \citenamefont {Kataria}, \citenamefont {Chandra},
  \citenamefont {Fallahi}, \citenamefont {Gardner},\ and\ \citenamefont
  {Manfra}}]{recent___2}%
  \BibitemOpen
  \bibfield  {author} {\bibinfo {author} {\bibfnamefont {A.}~\bibnamefont
  {Gupta}}, \bibinfo {author} {\bibfnamefont {J.~J.}\ \bibnamefont {Heremans}},
  \bibinfo {author} {\bibfnamefont {G.}~\bibnamefont {Kataria}}, \bibinfo
  {author} {\bibfnamefont {M.}~\bibnamefont {Chandra}}, \bibinfo {author}
  {\bibfnamefont {S.}~\bibnamefont {Fallahi}}, \bibinfo {author} {\bibfnamefont
  {G.~C.}\ \bibnamefont {Gardner}},\ and\ \bibinfo {author} {\bibfnamefont
  {M.~J.}\ \bibnamefont {Manfra}},\ }\href
  {https://doi.org/10.1103/PhysRevLett.126.076803} {\bibfield  {journal}
  {\bibinfo  {journal} {Phys. Rev. Lett.}\ }\textbf {\bibinfo {volume} {126}},\
  \bibinfo {pages} {076803} (\bibinfo {year} {2021})}\BibitemShut {NoStop}%
\bibitem [{\citenamefont {Scaffidi}\ \emph {et~al.}(2017)\citenamefont
  {Scaffidi}, \citenamefont {Nandi}, \citenamefont {Schmidt}, \citenamefont
  {Mackenzie},\ and\ \citenamefont {Moore}}]{ph_tr_num}%
  \BibitemOpen
  \bibfield  {author} {\bibinfo {author} {\bibfnamefont {T.}~\bibnamefont
  {Scaffidi}}, \bibinfo {author} {\bibfnamefont {N.}~\bibnamefont {Nandi}},
  \bibinfo {author} {\bibfnamefont {B.}~\bibnamefont {Schmidt}}, \bibinfo
  {author} {\bibfnamefont {A.~P.}\ \bibnamefont {Mackenzie}},\ and\ \bibinfo
  {author} {\bibfnamefont {J.~E.}\ \bibnamefont {Moore}},\ }\href
  {https://doi.org/10.1103/PhysRevLett.118.226601} {\bibfield  {journal}
  {\bibinfo  {journal} {Phys. Rev. Lett.}\ }\textbf {\bibinfo {volume} {118}},\
  \bibinfo {pages} {226601} (\bibinfo {year} {2017})}\BibitemShut {NoStop}%
\bibitem [{\citenamefont {Guo}\ \emph {et~al.}(2017)\citenamefont {Guo},
  \citenamefont {Ilseven}, \citenamefont {Falkovich},\ and\ \citenamefont
  {Levitov}}]{Levitov_et_al_2}%
  \BibitemOpen
  \bibfield  {author} {\bibinfo {author} {\bibfnamefont {H.}~\bibnamefont
  {Guo}}, \bibinfo {author} {\bibfnamefont {E.}~\bibnamefont {Ilseven}},
  \bibinfo {author} {\bibfnamefont {G.}~\bibnamefont {Falkovich}},\ and\
  \bibinfo {author} {\bibfnamefont {L.~S.}\ \bibnamefont {Levitov}},\ }\href
  {https://doi.org/10.1073/pnas.1612181114} {\bibfield  {journal} {\bibinfo
  {journal} {PNAS}\ }\textbf {\bibinfo {volume} {114}},\ \bibinfo {pages}
  {3068} (\bibinfo {year} {2017})}\BibitemShut {NoStop}%
\bibitem [{\citenamefont {Lucas}(2017)}]{Lucas}%
  \BibitemOpen
  \bibfield  {author} {\bibinfo {author} {\bibfnamefont {A.}~\bibnamefont
  {Lucas}},\ }\href {https://doi.org/10.1103/PhysRevB.95.115425} {\bibfield
  {journal} {\bibinfo  {journal} {Phys. Rev. B}\ }\textbf {\bibinfo {volume}
  {95}},\ \bibinfo {pages} {115425} (\bibinfo {year} {2017})}\BibitemShut
  {NoStop}%
\bibitem [{\citenamefont {Pellegrino}\ \emph {et~al.}(2017)\citenamefont
  {Pellegrino}, \citenamefont {Torre},\ and\ \citenamefont {Polini}}]{eta_xy}%
  \BibitemOpen
  \bibfield  {author} {\bibinfo {author} {\bibfnamefont {F.~M.~D.}\
  \bibnamefont {Pellegrino}}, \bibinfo {author} {\bibfnamefont
  {I.}~\bibnamefont {Torre}},\ and\ \bibinfo {author} {\bibfnamefont
  {M.}~\bibnamefont {Polini}},\ }\href
  {https://doi.org/10.1103/PhysRevB.96.195401} {\bibfield  {journal} {\bibinfo
  {journal} {Phys. Rev. B}\ }\textbf {\bibinfo {volume} {96}},\ \bibinfo
  {pages} {195401} (\bibinfo {year} {2017})}\BibitemShut {NoStop}%
\bibitem [{\citenamefont {Alekseev}\ \emph
  {et~al.}(2017{\natexlab{a}})\citenamefont {Alekseev}, \citenamefont {Gornyi},
  \citenamefont {Dmitriev}, \citenamefont {Kachorovskii},\ and\ \citenamefont
  {Semina}}]{we_3}%
  \BibitemOpen
  \bibfield  {author} {\bibinfo {author} {\bibfnamefont {P.~S.}\ \bibnamefont
  {Alekseev}}, \bibinfo {author} {\bibfnamefont {I.~V.}\ \bibnamefont
  {Gornyi}}, \bibinfo {author} {\bibfnamefont {A.~P.}\ \bibnamefont
  {Dmitriev}}, \bibinfo {author} {\bibfnamefont {V.~Y.}\ \bibnamefont
  {Kachorovskii}},\ and\ \bibinfo {author} {\bibfnamefont {M.~A.}\ \bibnamefont
  {Semina}},\ }\href {https://doi.org/10.1134/S1063782617060033} {\bibfield
  {journal} {\bibinfo  {journal} {Semiconductors}\ }\textbf {\bibinfo {volume}
  {51}},\ \bibinfo {pages} {766} (\bibinfo {year}
  {2017}{\natexlab{a}})}\BibitemShut {NoStop}%
\bibitem [{\citenamefont {Lucas}\ and\ \citenamefont {Fong}(2018)}]{Lucas_2}%
  \BibitemOpen
  \bibfield  {author} {\bibinfo {author} {\bibfnamefont {A.}~\bibnamefont
  {Lucas}}\ and\ \bibinfo {author} {\bibfnamefont {K.~C.}\ \bibnamefont
  {Fong}},\ }\href {https://doi.org/10.1088/1361-648x/aaa274} {\bibfield
  {journal} {\bibinfo  {journal} {Journal of Physics: Condensed Matter}\
  }\textbf {\bibinfo {volume} {30}},\ \bibinfo {pages} {053001} (\bibinfo
  {year} {2018})}\BibitemShut {NoStop}%
\bibitem [{\citenamefont {Alekseev}\ \emph
  {et~al.}(2017{\natexlab{b}})\citenamefont {Alekseev}, \citenamefont
  {Dmitriev}, \citenamefont {Gornyi}, \citenamefont {Kachorovskii},
  \citenamefont {Narozhny}, \citenamefont {Sch\"utt},\ and\ \citenamefont
  {Titov}}]{we_4}%
  \BibitemOpen
  \bibfield  {author} {\bibinfo {author} {\bibfnamefont {P.~S.}\ \bibnamefont
  {Alekseev}}, \bibinfo {author} {\bibfnamefont {A.~P.}\ \bibnamefont
  {Dmitriev}}, \bibinfo {author} {\bibfnamefont {I.~V.}\ \bibnamefont
  {Gornyi}}, \bibinfo {author} {\bibfnamefont {V.~Y.}\ \bibnamefont
  {Kachorovskii}}, \bibinfo {author} {\bibfnamefont {B.~N.}\ \bibnamefont
  {Narozhny}}, \bibinfo {author} {\bibfnamefont {M.}~\bibnamefont {Sch\"utt}},\
  and\ \bibinfo {author} {\bibfnamefont {M.}~\bibnamefont {Titov}},\ }\href
  {https://doi.org/10.1103/PhysRevB.95.165410} {\bibfield  {journal} {\bibinfo
  {journal} {Phys. Rev. B}\ }\textbf {\bibinfo {volume} {95}},\ \bibinfo
  {pages} {165410} (\bibinfo {year} {2017}{\natexlab{b}})}\BibitemShut
  {NoStop}%
\bibitem [{\citenamefont {Alekseev}\ \emph
  {et~al.}(2018{\natexlab{a}})\citenamefont {Alekseev}, \citenamefont
  {Dmitriev}, \citenamefont {Gornyi}, \citenamefont {Kachorovskii},
  \citenamefont {Narozhny},\ and\ \citenamefont {Titov}}]{we_5_1}%
  \BibitemOpen
  \bibfield  {author} {\bibinfo {author} {\bibfnamefont {P.~S.}\ \bibnamefont
  {Alekseev}}, \bibinfo {author} {\bibfnamefont {A.~P.}\ \bibnamefont
  {Dmitriev}}, \bibinfo {author} {\bibfnamefont {I.~V.}\ \bibnamefont
  {Gornyi}}, \bibinfo {author} {\bibfnamefont {V.~Y.}\ \bibnamefont
  {Kachorovskii}}, \bibinfo {author} {\bibfnamefont {B.~N.}\ \bibnamefont
  {Narozhny}},\ and\ \bibinfo {author} {\bibfnamefont {M.}~\bibnamefont
  {Titov}},\ }\href {https://doi.org/10.1103/PhysRevB.97.085109} {\bibfield
  {journal} {\bibinfo  {journal} {Phys. Rev. B}\ }\textbf {\bibinfo {volume}
  {97}},\ \bibinfo {pages} {085109} (\bibinfo {year}
  {2018}{\natexlab{a}})}\BibitemShut {NoStop}%
\bibitem [{\citenamefont {Alekseev}\ \emph
  {et~al.}(2018{\natexlab{b}})\citenamefont {Alekseev}, \citenamefont
  {Dmitriev}, \citenamefont {Gornyi}, \citenamefont {Kachorovskii},
  \citenamefont {Narozhny},\ and\ \citenamefont {Titov}}]{we_5_2}%
  \BibitemOpen
  \bibfield  {author} {\bibinfo {author} {\bibfnamefont {P.~S.}\ \bibnamefont
  {Alekseev}}, \bibinfo {author} {\bibfnamefont {A.~P.}\ \bibnamefont
  {Dmitriev}}, \bibinfo {author} {\bibfnamefont {I.~V.}\ \bibnamefont
  {Gornyi}}, \bibinfo {author} {\bibfnamefont {V.~Y.}\ \bibnamefont
  {Kachorovskii}}, \bibinfo {author} {\bibfnamefont {B.~N.}\ \bibnamefont
  {Narozhny}},\ and\ \bibinfo {author} {\bibfnamefont {M.}~\bibnamefont
  {Titov}},\ }\href {https://doi.org/10.1103/PhysRevB.98.125111} {\bibfield
  {journal} {\bibinfo  {journal} {Phys. Rev. B}\ }\textbf {\bibinfo {volume}
  {98}},\ \bibinfo {pages} {125111} (\bibinfo {year}
  {2018}{\natexlab{b}})}\BibitemShut {NoStop}%
\bibitem [{\citenamefont {Alekseev}\ and\ \citenamefont {Semina}(2018)}]{we_5}%
  \BibitemOpen
  \bibfield  {author} {\bibinfo {author} {\bibfnamefont {P.~S.}\ \bibnamefont
  {Alekseev}}\ and\ \bibinfo {author} {\bibfnamefont {M.~A.}\ \bibnamefont
  {Semina}},\ }\href {https://doi.org/10.1103/PhysRevB.98.165412} {\bibfield
  {journal} {\bibinfo  {journal} {Phys. Rev. B}\ }\textbf {\bibinfo {volume}
  {98}},\ \bibinfo {pages} {165412} (\bibinfo {year} {2018})}\BibitemShut
  {NoStop}%
\bibitem [{\citenamefont {Alekseev}\ and\ \citenamefont {Semina}(2019)}]{we_6}%
  \BibitemOpen
  \bibfield  {author} {\bibinfo {author} {\bibfnamefont {P.~S.}\ \bibnamefont
  {Alekseev}}\ and\ \bibinfo {author} {\bibfnamefont {M.~A.}\ \bibnamefont
  {Semina}},\ }\href {https://doi.org/10.1103/PhysRevB.100.125419} {\bibfield
  {journal} {\bibinfo  {journal} {Phys. Rev. B}\ }\textbf {\bibinfo {volume}
  {100}},\ \bibinfo {pages} {125419} (\bibinfo {year} {2019})}\BibitemShut
  {NoStop}%
\bibitem [{\citenamefont {Kashuba}\ \emph {et~al.}(2018)\citenamefont
  {Kashuba}, \citenamefont {Trauzettel},\ and\ \citenamefont
  {Molenkamp}}]{recentest_}%
  \BibitemOpen
  \bibfield  {author} {\bibinfo {author} {\bibfnamefont {O.}~\bibnamefont
  {Kashuba}}, \bibinfo {author} {\bibfnamefont {B.}~\bibnamefont
  {Trauzettel}},\ and\ \bibinfo {author} {\bibfnamefont {L.~W.}\ \bibnamefont
  {Molenkamp}},\ }\href {https://doi.org/10.1103/PhysRevB.97.205129} {\bibfield
   {journal} {\bibinfo  {journal} {Phys. Rev. B}\ }\textbf {\bibinfo {volume}
  {97}},\ \bibinfo {pages} {205129} (\bibinfo {year} {2018})}\BibitemShut
  {NoStop}%
\bibitem [{\citenamefont {Moessner}\ \emph {et~al.}(2018)\citenamefont
  {Moessner}, \citenamefont {Sur\'owka},\ and\ \citenamefont
  {Witkowski}}]{recentest___breathing_flow}%
  \BibitemOpen
  \bibfield  {author} {\bibinfo {author} {\bibfnamefont {R.}~\bibnamefont
  {Moessner}}, \bibinfo {author} {\bibfnamefont {P.}~\bibnamefont
  {Sur\'owka}},\ and\ \bibinfo {author} {\bibfnamefont {P.}~\bibnamefont
  {Witkowski}},\ }\href {https://doi.org/10.1103/PhysRevB.97.161112} {\bibfield
   {journal} {\bibinfo  {journal} {Phys. Rev. B}\ }\textbf {\bibinfo {volume}
  {97}},\ \bibinfo {pages} {161112(R)} (\bibinfo {year} {2018})}\BibitemShut
  {NoStop}%
\bibitem [{\citenamefont {Semenyakin}\ and\ \citenamefont
  {Falkovich}(2018)}]{recentest2}%
  \BibitemOpen
  \bibfield  {author} {\bibinfo {author} {\bibfnamefont {M.}~\bibnamefont
  {Semenyakin}}\ and\ \bibinfo {author} {\bibfnamefont {G.}~\bibnamefont
  {Falkovich}},\ }\href {https://doi.org/10.1103/PhysRevB.97.085127} {\bibfield
   {journal} {\bibinfo  {journal} {Phys. Rev. B}\ }\textbf {\bibinfo {volume}
  {97}},\ \bibinfo {pages} {085127} (\bibinfo {year} {2018})}\BibitemShut
  {NoStop}%
\bibitem [{\citenamefont {Lucas}\ and\ \citenamefont
  {Das~Sarma}(2018)}]{L_n_1}%
  \BibitemOpen
  \bibfield  {author} {\bibinfo {author} {\bibfnamefont {A.}~\bibnamefont
  {Lucas}}\ and\ \bibinfo {author} {\bibfnamefont {S.}~\bibnamefont
  {Das~Sarma}},\ }\href {https://doi.org/10.1103/PhysRevB.97.115449} {\bibfield
   {journal} {\bibinfo  {journal} {Phys. Rev. B}\ }\textbf {\bibinfo {volume}
  {97}},\ \bibinfo {pages} {115449} (\bibinfo {year} {2018})}\BibitemShut
  {NoStop}%
\bibitem [{\citenamefont {Cohen}\ and\ \citenamefont
  {Goldstein}(2018)}]{recentest3}%
  \BibitemOpen
  \bibfield  {author} {\bibinfo {author} {\bibfnamefont {R.}~\bibnamefont
  {Cohen}}\ and\ \bibinfo {author} {\bibfnamefont {M.}~\bibnamefont
  {Goldstein}},\ }\href {https://doi.org/10.1103/PhysRevB.98.235103} {\bibfield
   {journal} {\bibinfo  {journal} {Phys. Rev. B}\ }\textbf {\bibinfo {volume}
  {98}},\ \bibinfo {pages} {235103} (\bibinfo {year} {2018})}\BibitemShut
  {NoStop}%
\bibitem [{\citenamefont {Alekseev}(2018)}]{vis_res}%
  \BibitemOpen
  \bibfield  {author} {\bibinfo {author} {\bibfnamefont {P.~S.}\ \bibnamefont
  {Alekseev}},\ }\href {https://doi.org/10.1103/PhysRevB.98.165440} {\bibfield
  {journal} {\bibinfo  {journal} {Phys. Rev. B}\ }\textbf {\bibinfo {volume}
  {98}},\ \bibinfo {pages} {165440} (\bibinfo {year} {2018})}\BibitemShut
  {NoStop}%
\bibitem [{\citenamefont {Holder}\ \emph {et~al.}(2019)\citenamefont {Holder},
  \citenamefont {Queiroz}, \citenamefont {Scaffidi}, \citenamefont
  {Silberstein}, \citenamefont {Rozen}, \citenamefont {Sulpizio}, \citenamefont
  {Ella}, \citenamefont {Ilani},\ and\ \citenamefont
  {Stern}}]{ph_tr_num__ball_formulas}%
  \BibitemOpen
  \bibfield  {author} {\bibinfo {author} {\bibfnamefont {T.}~\bibnamefont
  {Holder}}, \bibinfo {author} {\bibfnamefont {R.}~\bibnamefont {Queiroz}},
  \bibinfo {author} {\bibfnamefont {T.}~\bibnamefont {Scaffidi}}, \bibinfo
  {author} {\bibfnamefont {N.}~\bibnamefont {Silberstein}}, \bibinfo {author}
  {\bibfnamefont {A.}~\bibnamefont {Rozen}}, \bibinfo {author} {\bibfnamefont
  {J.~A.}\ \bibnamefont {Sulpizio}}, \bibinfo {author} {\bibfnamefont
  {L.}~\bibnamefont {Ella}}, \bibinfo {author} {\bibfnamefont {S.}~\bibnamefont
  {Ilani}},\ and\ \bibinfo {author} {\bibfnamefont {A.}~\bibnamefont {Stern}},\
  }\href {https://doi.org/10.1103/PhysRevB.100.245305} {\bibfield  {journal}
  {\bibinfo  {journal} {Phys. Rev. B}\ }\textbf {\bibinfo {volume} {100}},\
  \bibinfo {pages} {245305} (\bibinfo {year} {2019})}\BibitemShut {NoStop}%
\bibitem [{\citenamefont {Khoo}\ and\ \citenamefont
  {Villadiego}(2019)}]{Khoo_Villadiego}%
  \BibitemOpen
  \bibfield  {author} {\bibinfo {author} {\bibfnamefont {J.~Y.}\ \bibnamefont
  {Khoo}}\ and\ \bibinfo {author} {\bibfnamefont {I.~S.}\ \bibnamefont
  {Villadiego}},\ }\href {https://doi.org/10.1103/PhysRevB.99.075434}
  {\bibfield  {journal} {\bibinfo  {journal} {Phys. Rev. B}\ }\textbf {\bibinfo
  {volume} {99}},\ \bibinfo {pages} {075434} (\bibinfo {year}
  {2019})}\BibitemShut {NoStop}%
\bibitem [{\citenamefont {Alekseev}(2019)}]{future}%
  \BibitemOpen
  \bibfield  {author} {\bibinfo {author} {\bibfnamefont {P.~S.}\ \bibnamefont
  {Alekseev}},\ }\href {https://doi.org/10.1134/S1063782619100026} {\bibfield
  {journal} {\bibinfo  {journal} {Semiconductors}\ }\textbf {\bibinfo {volume}
  {53}},\ \bibinfo {pages} {1367} (\bibinfo {year} {2019})}\BibitemShut
  {NoStop}%
\bibitem [{\citenamefont {Alekseev}\ and\ \citenamefont
  {Alekseeva}(2019)}]{Alekseev_Alekseeva}%
  \BibitemOpen
  \bibfield  {author} {\bibinfo {author} {\bibfnamefont {P.~S.}\ \bibnamefont
  {Alekseev}}\ and\ \bibinfo {author} {\bibfnamefont {A.~P.}\ \bibnamefont
  {Alekseeva}},\ }\href {https://doi.org/10.1103/PhysRevLett.123.236801}
  {\bibfield  {journal} {\bibinfo  {journal} {Phys. Rev. Lett.}\ }\textbf
  {\bibinfo {volume} {123}},\ \bibinfo {pages} {236801} (\bibinfo {year}
  {2019})}\BibitemShut {NoStop}%
\bibitem [{\citenamefont {Alekseev}\ and\ \citenamefont
  {Dmitriev}(2020)}]{Alekseev_Dmitriev}%
  \BibitemOpen
  \bibfield  {author} {\bibinfo {author} {\bibfnamefont {P.~S.}\ \bibnamefont
  {Alekseev}}\ and\ \bibinfo {author} {\bibfnamefont {A.~P.}\ \bibnamefont
  {Dmitriev}},\ }\href {https://doi.org/10.1103/PhysRevB.102.241409} {\bibfield
   {journal} {\bibinfo  {journal} {Phys. Rev. B}\ }\textbf {\bibinfo {volume}
  {102}},\ \bibinfo {pages} {241409(R)} (\bibinfo {year} {2020})}\BibitemShut
  {NoStop}%
\bibitem [{\citenamefont {Trachenko}\ and\ \citenamefont
  {Brazhkin}(2020)}]{future2}%
  \BibitemOpen
  \bibfield  {author} {\bibinfo {author} {\bibfnamefont {K.}~\bibnamefont
  {Trachenko}}\ and\ \bibinfo {author} {\bibfnamefont {V.~V.}\ \bibnamefont
  {Brazhkin}},\ }\href {https://doi.org/10.1126/sciadv.aba3747} {\bibfield
  {journal} {\bibinfo  {journal} {Sci. Adv.}\ }\textbf {\bibinfo {volume}
  {6}},\ \bibinfo {pages} {eaba3747} (\bibinfo {year} {2020})}\BibitemShut
  {NoStop}%
\bibitem [{\citenamefont {Afanasiev}\ \emph
  {et~al.}(2021{\natexlab{a}})\citenamefont {Afanasiev}, \citenamefont
  {Alekseev}, \citenamefont {Greshnov},\ and\ \citenamefont {Semina}}]{we_7}%
  \BibitemOpen
  \bibfield  {author} {\bibinfo {author} {\bibfnamefont {A.~N.}\ \bibnamefont
  {Afanasiev}}, \bibinfo {author} {\bibfnamefont {P.~S.}\ \bibnamefont
  {Alekseev}}, \bibinfo {author} {\bibfnamefont {A.~A.}\ \bibnamefont
  {Greshnov}},\ and\ \bibinfo {author} {\bibfnamefont {M.~A.}\ \bibnamefont
  {Semina}},\ }\href {https://doi.org/10.1134/S1063782621070022} {\bibfield
  {journal} {\bibinfo  {journal} {Semiconductors}\ }\textbf {\bibinfo {volume}
  {55}},\ \bibinfo {pages} {562} (\bibinfo {year}
  {2021}{\natexlab{a}})}\BibitemShut {NoStop}%
\bibitem [{\citenamefont {Afanasiev}\ \emph
  {et~al.}(2021{\natexlab{b}})\citenamefont {Afanasiev}, \citenamefont
  {Alekseev}, \citenamefont {Greshnov},\ and\ \citenamefont
  {Semina}}]{we_8___ph_tr_formulas}%
  \BibitemOpen
  \bibfield  {author} {\bibinfo {author} {\bibfnamefont {A.~N.}\ \bibnamefont
  {Afanasiev}}, \bibinfo {author} {\bibfnamefont {P.~S.}\ \bibnamefont
  {Alekseev}}, \bibinfo {author} {\bibfnamefont {A.~A.}\ \bibnamefont
  {Greshnov}},\ and\ \bibinfo {author} {\bibfnamefont {M.~A.}\ \bibnamefont
  {Semina}},\ }\href {https://doi.org/10.1103/PhysRevB.104.195415} {\bibfield
  {journal} {\bibinfo  {journal} {Phys. Rev. B}\ }\textbf {\bibinfo {volume}
  {104}},\ \bibinfo {pages} {195415} (\bibinfo {year}
  {2021}{\natexlab{b}})}\BibitemShut {NoStop}%
\bibitem [{\citenamefont {Dyakonov}\ and\ \citenamefont
  {Perel}(1971{\natexlab{a}})}]{Dyakonov_Perel_1}%
  \BibitemOpen
  \bibfield  {author} {\bibinfo {author} {\bibfnamefont {M.~I.}\ \bibnamefont
  {Dyakonov}}\ and\ \bibinfo {author} {\bibfnamefont {V.~I.}\ \bibnamefont
  {Perel}},\ }\href
  {http://jetpletters.ru/cgi-bin/articles/download.cgi/1587/article_24366.pdf}
  {\bibfield  {journal} {\bibinfo  {journal} {Sov. Phys. JETP Lett.}\ }\textbf
  {\bibinfo {volume} {13}},\ \bibinfo {pages} {467} (\bibinfo {year}
  {1971}{\natexlab{a}})}\BibitemShut {NoStop}%
\bibitem [{\citenamefont {Dyakonov}\ and\ \citenamefont
  {Perel}(1971{\natexlab{b}})}]{Dyakonov_Perel_2}%
  \BibitemOpen
  \bibfield  {author} {\bibinfo {author} {\bibfnamefont {M.~I.}\ \bibnamefont
  {Dyakonov}}\ and\ \bibinfo {author} {\bibfnamefont {V.~I.}\ \bibnamefont
  {Perel}},\ }\href
  {https://doi.org/https://doi.org/10.1016/0375-9601(71)90196-4} {\bibfield
  {journal} {\bibinfo  {journal} {Physics Letters A}\ }\textbf {\bibinfo
  {volume} {35}},\ \bibinfo {pages} {459} (\bibinfo {year}
  {1971}{\natexlab{b}})}\BibitemShut {NoStop}%
\bibitem [{\citenamefont {Hirsch}(1999)}]{Hirsch}%
  \BibitemOpen
  \bibfield  {author} {\bibinfo {author} {\bibfnamefont {J.~E.}\ \bibnamefont
  {Hirsch}},\ }\href {https://doi.org/10.1103/PhysRevLett.83.1834} {\bibfield
  {journal} {\bibinfo  {journal} {Phys. Rev. Lett.}\ }\textbf {\bibinfo
  {volume} {83}},\ \bibinfo {pages} {1834} (\bibinfo {year}
  {1999})}\BibitemShut {NoStop}%
\bibitem [{\citenamefont {Dyakonov}\ and\ \citenamefont
  {Khaetskii}(2017)}]{book}%
  \BibitemOpen
  \bibfield  {author} {\bibinfo {author} {\bibfnamefont {M.}~\bibnamefont
  {Dyakonov}}\ and\ \bibinfo {author} {\bibfnamefont {A.}~\bibnamefont
  {Khaetskii}},\ }\href {https://doi.org/10.1007/978-3-319-65436-2} {\emph
  {\bibinfo {title} {Spin {P}hysics in {S}emiconductors}}},\ edited by\
  \bibinfo {editor} {\bibfnamefont {M.}~\bibnamefont {Dyakonov}}\ (\bibinfo
  {publisher} {Springer},\ \bibinfo {year} {2017})\BibitemShut {NoStop}%
\bibitem [{\citenamefont {Dyakonov}(2007)}]{Dyakonov__spin_Hall_magnetores}%
  \BibitemOpen
  \bibfield  {author} {\bibinfo {author} {\bibfnamefont {M.~I.}\ \bibnamefont
  {Dyakonov}},\ }\href {https://doi.org/10.1103/PhysRevLett.99.126601}
  {\bibfield  {journal} {\bibinfo  {journal} {Phys. Rev. Lett.}\ }\textbf
  {\bibinfo {volume} {99}},\ \bibinfo {pages} {126601} (\bibinfo {year}
  {2007})}\BibitemShut {NoStop}%
\bibitem [{\citenamefont {Alekseev}\ and\ \citenamefont
  {Dyakonov}(2019)}]{Alekseev_Dyakonov__spin_Hall_magnetoimped}%
  \BibitemOpen
  \bibfield  {author} {\bibinfo {author} {\bibfnamefont {P.~S.}\ \bibnamefont
  {Alekseev}}\ and\ \bibinfo {author} {\bibfnamefont {M.~I.}\ \bibnamefont
  {Dyakonov}},\ }\href {https://doi.org/10.1103/PhysRevB.100.081301} {\bibfield
   {journal} {\bibinfo  {journal} {Phys. Rev. B}\ }\textbf {\bibinfo {volume}
  {100}},\ \bibinfo {pages} {081301(R)} (\bibinfo {year} {2019})}\BibitemShut
  {NoStop}%
\bibitem [{\citenamefont {Matsuo}\ \emph {et~al.}(2017)\citenamefont {Matsuo},
  \citenamefont {Ohnuma},\ and\ \citenamefont
  {Maekawa}}]{Maekawa_et_al__rotat_visc_1}%
  \BibitemOpen
  \bibfield  {author} {\bibinfo {author} {\bibfnamefont {M.}~\bibnamefont
  {Matsuo}}, \bibinfo {author} {\bibfnamefont {Y.}~\bibnamefont {Ohnuma}},\
  and\ \bibinfo {author} {\bibfnamefont {S.}~\bibnamefont {Maekawa}},\ }\href
  {https://doi.org/10.1103/PhysRevB.96.020401} {\bibfield  {journal} {\bibinfo
  {journal} {Phys. Rev. B}\ }\textbf {\bibinfo {volume} {96}},\ \bibinfo
  {pages} {020401(R)} (\bibinfo {year} {2017})}\BibitemShut {NoStop}%
\bibitem [{\citenamefont {Matsuo}\ \emph {et~al.}(2020)\citenamefont {Matsuo},
  \citenamefont {Bandurin}, \citenamefont {Ohnuma}, \citenamefont {Tsutsumi},\
  and\ \citenamefont {Maekawa}}]{Maekawa_et_al__rotat_visc_2}%
  \BibitemOpen
  \bibfield  {author} {\bibinfo {author} {\bibfnamefont {M.}~\bibnamefont
  {Matsuo}}, \bibinfo {author} {\bibfnamefont {D.~A.}\ \bibnamefont
  {Bandurin}}, \bibinfo {author} {\bibfnamefont {Y.}~\bibnamefont {Ohnuma}},
  \bibinfo {author} {\bibfnamefont {Y.}~\bibnamefont {Tsutsumi}},\ and\
  \bibinfo {author} {\bibfnamefont {S.}~\bibnamefont {Maekawa}},\ }\href@noop
  {} {} (\bibinfo {year} {2020}),\ \Eprint {https://arxiv.org/abs/2005.01493}
  {arXiv:2005.01493 [cond-mat.mes-hall]} \BibitemShut {NoStop}%
\bibitem [{\citenamefont {Takahashi}\ \emph {et~al.}(2016)\citenamefont
  {Takahashi}, \citenamefont {Matsuo}, \citenamefont {Ono}, \citenamefont
  {Harii}, \citenamefont {Chudo}, \citenamefont {Okayasu}, \citenamefont
  {Ieda}, \citenamefont {Takahashi}, \citenamefont {Maekawa},\ and\
  \citenamefont {Saitoh}}]{Maekawa_et_al__rotat_visc_3}%
  \BibitemOpen
  \bibfield  {author} {\bibinfo {author} {\bibfnamefont {R.}~\bibnamefont
  {Takahashi}}, \bibinfo {author} {\bibfnamefont {M.}~\bibnamefont {Matsuo}},
  \bibinfo {author} {\bibfnamefont {M.}~\bibnamefont {Ono}}, \bibinfo {author}
  {\bibfnamefont {K.}~\bibnamefont {Harii}}, \bibinfo {author} {\bibfnamefont
  {H.}~\bibnamefont {Chudo}}, \bibinfo {author} {\bibfnamefont
  {S.}~\bibnamefont {Okayasu}}, \bibinfo {author} {\bibfnamefont
  {J.}~\bibnamefont {Ieda}}, \bibinfo {author} {\bibfnamefont {S.}~\bibnamefont
  {Takahashi}}, \bibinfo {author} {\bibfnamefont {S.}~\bibnamefont {Maekawa}},\
  and\ \bibinfo {author} {\bibfnamefont {E.}~\bibnamefont {Saitoh}},\ }\href
  {https://doi.org/10.1038/nphys3526} {\bibfield  {journal} {\bibinfo
  {journal} {Nature Physics}\ }\textbf {\bibinfo {volume} {12}},\ \bibinfo
  {pages} {52} (\bibinfo {year} {2016})}\BibitemShut {NoStop}%
\bibitem [{\citenamefont {Takahashi}\ \emph {et~al.}(2020)\citenamefont
  {Takahashi}, \citenamefont {Chudo}, \citenamefont {Matsuo}, \citenamefont
  {Harii}, \citenamefont {Ohnuma}, \citenamefont {Maekawa},\ and\ \citenamefont
  {Saitoh}}]{Maekawa_et_al__rotat_visc_4}%
  \BibitemOpen
  \bibfield  {author} {\bibinfo {author} {\bibfnamefont {R.}~\bibnamefont
  {Takahashi}}, \bibinfo {author} {\bibfnamefont {H.}~\bibnamefont {Chudo}},
  \bibinfo {author} {\bibfnamefont {M.}~\bibnamefont {Matsuo}}, \bibinfo
  {author} {\bibfnamefont {K.}~\bibnamefont {Harii}}, \bibinfo {author}
  {\bibfnamefont {Y.}~\bibnamefont {Ohnuma}}, \bibinfo {author} {\bibfnamefont
  {S.}~\bibnamefont {Maekawa}},\ and\ \bibinfo {author} {\bibfnamefont
  {E.}~\bibnamefont {Saitoh}},\ }\href
  {https://doi.org/10.1038/s41467-020-16753-0} {\bibfield  {journal} {\bibinfo
  {journal} {Nature Communications}\ }\textbf {\bibinfo {volume} {11}},\
  \bibinfo {pages} {3009} (\bibinfo {year} {2020})}\BibitemShut {NoStop}%
\bibitem [{\citenamefont {Doornenbal}\ \emph {et~al.}(2019)\citenamefont
  {Doornenbal}, \citenamefont {Polini},\ and\ \citenamefont
  {Duine}}]{Polini__rotat_visc}%
  \BibitemOpen
  \bibfield  {author} {\bibinfo {author} {\bibfnamefont {R.~J.}\ \bibnamefont
  {Doornenbal}}, \bibinfo {author} {\bibfnamefont {M.}~\bibnamefont {Polini}},\
  and\ \bibinfo {author} {\bibfnamefont {R.~A.}\ \bibnamefont {Duine}},\ }\href
  {https://doi.org/10.1088/2515-7639/aaf8fb} {\bibfield  {journal} {\bibinfo
  {journal} {Journal of Physics: Materials}\ }\textbf {\bibinfo {volume} {2}},\
  \bibinfo {pages} {015006} (\bibinfo {year} {2019})}\BibitemShut {NoStop}%
\bibitem [{\citenamefont {Glazov}\ and\ \citenamefont
  {Ivchenko}(2002)}]{Glazov_Ivchenko_1}%
  \BibitemOpen
  \bibfield  {author} {\bibinfo {author} {\bibfnamefont {M.~M.}\ \bibnamefont
  {Glazov}}\ and\ \bibinfo {author} {\bibfnamefont {E.~L.}\ \bibnamefont
  {Ivchenko}},\ }\href {https://doi.org/https://doi.org/10.1134/1.1490009}
  {\bibfield  {journal} {\bibinfo  {journal} {JETP Lett.}\ }\textbf {\bibinfo
  {volume} {75}},\ \bibinfo {pages} {403} (\bibinfo {year} {2002})}\BibitemShut
  {NoStop}%
\bibitem [{\citenamefont {Glazov}\ and\ \citenamefont
  {Ivchenko}(2004)}]{Glazov_Ivchenko_2}%
  \BibitemOpen
  \bibfield  {author} {\bibinfo {author} {\bibfnamefont {M.~M.}\ \bibnamefont
  {Glazov}}\ and\ \bibinfo {author} {\bibfnamefont {E.~L.}\ \bibnamefont
  {Ivchenko}},\ }\href {https://doi.org/10.1134/1.1854815} {\bibfield
  {journal} {\bibinfo  {journal} {JETP}\ }\textbf {\bibinfo {volume} {99}},\
  \bibinfo {pages} {1279} (\bibinfo {year} {2004})}\BibitemShut {NoStop}%
\bibitem [{\citenamefont {D'Amico}\ and\ \citenamefont
  {Vignale}(2000)}]{D_Amico_Vignale____coulomb_drag_1}%
  \BibitemOpen
  \bibfield  {author} {\bibinfo {author} {\bibfnamefont {I.}~\bibnamefont
  {D'Amico}}\ and\ \bibinfo {author} {\bibfnamefont {G.}~\bibnamefont
  {Vignale}},\ }\href {https://doi.org/10.1103/PhysRevB.62.4853} {\bibfield
  {journal} {\bibinfo  {journal} {Phys. Rev. B}\ }\textbf {\bibinfo {volume}
  {62}},\ \bibinfo {pages} {4853} (\bibinfo {year} {2000})}\BibitemShut
  {NoStop}%
\bibitem [{\citenamefont {D'Amico}\ and\ \citenamefont
  {Vignale}(2003)}]{D_Amico_Vignale____coulomb_drag_2}%
  \BibitemOpen
  \bibfield  {author} {\bibinfo {author} {\bibfnamefont {I.}~\bibnamefont
  {D'Amico}}\ and\ \bibinfo {author} {\bibfnamefont {G.}~\bibnamefont
  {Vignale}},\ }\href {https://doi.org/10.1103/PhysRevB.68.045307} {\bibfield
  {journal} {\bibinfo  {journal} {Phys. Rev. B}\ }\textbf {\bibinfo {volume}
  {68}},\ \bibinfo {pages} {045307} (\bibinfo {year} {2003})}\BibitemShut
  {NoStop}%
\bibitem [{\citenamefont {Dyakonov}\ and\ \citenamefont
  {Perel}(1972)}]{Dyakonov_Perel_3__spin_rel}%
  \BibitemOpen
  \bibfield  {author} {\bibinfo {author} {\bibfnamefont {M.~I.}\ \bibnamefont
  {Dyakonov}}\ and\ \bibinfo {author} {\bibfnamefont {V.~I.}\ \bibnamefont
  {Perel}},\ }\href@noop {} {\bibfield  {journal} {\bibinfo  {journal} {Soviet
  Physics Solid State}\ }\textbf {\bibinfo {volume} {13}},\ \bibinfo {pages}
  {3023} (\bibinfo {year} {1972})}\BibitemShut {NoStop}%
\bibitem [{\citenamefont {Poiseuille}(1840)}]{Poiseuille}%
  \BibitemOpen
  \bibfield  {author} {\bibinfo {author} {\bibfnamefont {J.~L.~M.}\
  \bibnamefont {Poiseuille}},\ }\href@noop {} {\bibfield  {journal} {\bibinfo
  {journal} {C. R. Acad. Sci.}\ }\textbf {\bibinfo {volume} {11}},\ \bibinfo
  {pages} {961} (\bibinfo {year} {1840})}\BibitemShut {NoStop}%
\bibitem [{\citenamefont {Womersley}(1955)}]{Womersley}%
  \BibitemOpen
  \bibfield  {author} {\bibinfo {author} {\bibfnamefont {J.~R.}\ \bibnamefont
  {Womersley}},\ }\href
  {https://doi.org/https://doi.org/10.1113/jphysiol.1955.sp005276} {\bibfield
  {journal} {\bibinfo  {journal} {The Journal of Physiology}\ }\textbf
  {\bibinfo {volume} {127}},\ \bibinfo {pages} {553} (\bibinfo {year}
  {1955})}\BibitemShut {NoStop}%
\bibitem [{\citenamefont {Glazov}(2021)}]{Glazov___spin_Hall_in_hydr}%
  \BibitemOpen
  \bibfield  {author} {\bibinfo {author} {\bibfnamefont {M.~M.}\ \bibnamefont
  {Glazov}},\ }\href {https://doi.org/10.1088/2053-1583/ac3e04} {\bibfield
  {journal} {\bibinfo  {journal} {2D Materials}\ }\textbf {\bibinfo {volume}
  {9}},\ \bibinfo {pages} {015027} (\bibinfo {year} {2021})}\BibitemShut
  {NoStop}%
\bibitem [{\citenamefont {De~Groot}\ and\ \citenamefont
  {Mazur}(1962)}]{DeGroot1962}%
  \BibitemOpen
  \bibfield  {author} {\bibinfo {author} {\bibfnamefont {S.~R.}\ \bibnamefont
  {De~Groot}}\ and\ \bibinfo {author} {\bibfnamefont {P.}~\bibnamefont
  {Mazur}},\ }\href
  {https://books.google.ru/books?id=mfFyG9jfaMYC&dq=De+Groot+S+Rand+Mazur+P1962+Non-equilibrium+thermodynamics+(Amsterdam:North-Holland)&lr=&hl=ru&source=gbs_navlinks_s}
  {\emph {\bibinfo {title} {Non-Equilibrium Thermodynamics}}}\ (\bibinfo
  {publisher} {North-Holland},\ \bibinfo {address} {Amsterdam},\ \bibinfo
  {year} {1962})\BibitemShut {NoStop}%
\bibitem [{\citenamefont {Snider}\ and\ \citenamefont
  {Lewchuk}(1967)}]{Snider1967}%
  \BibitemOpen
  \bibfield  {author} {\bibinfo {author} {\bibfnamefont {R.~F.}\ \bibnamefont
  {Snider}}\ and\ \bibinfo {author} {\bibfnamefont {K.~S.}\ \bibnamefont
  {Lewchuk}},\ }\href {https://doi.org/10.1063/1.1841187} {\bibfield  {journal}
  {\bibinfo  {journal} {J. Chem. Phys.}\ }\textbf {\bibinfo {volume} {46}},\
  \bibinfo {pages} {3163} (\bibinfo {year} {1967})}\BibitemShut {NoStop}%
\end{thebibliography}%

\end{document}